%% file: driver.tex
\documentclass[11pt,a4paper]{article}
\usepackage{helvet}
\usepackage{graphicx}
\usepackage{color}
\usepackage{rotating}
\usepackage{eurosym}
\usepackage{wrapfig}
\usepackage{setspace}
\usepackage{eso-pic}
\usepackage{subfig}
\usepackage{afterpage}
\usepackage{sidecap}
\usepackage[numbers]{natbib}
\usepackage{aas_macros}
\usepackage{hyperref}
\usepackage{multirow}

\def\lsim{\raise0.3ex\hbox{$<$\kern-0.75em\raise-1.1ex\hbox{$\sim$}}}
\def\gsim{\raise0.3ex\hbox{$>$\kern-0.75em\raise-1.1ex\hbox{$\sim$}}}

\def\beq{\begin{equation}} \def\eeq{\end{equation}} \def\bea{\begin{eqnarray}}
\def\eea{\end{eqnarray}}  
\def\be{\begin{equation}} \def\ee{\end{equation}} \def\ba{\begin{eqnarray}}
\def\ea{\end{eqnarray}} 
\def\dalemb#1#2{{\vbox{\hrule height.#2pt \hbox{\vrule width.#2pt height#1pt
\kern#1pt \vrule width.#2pt} \hrule height.#2pt}}}

\def\ba{\begin{eqnarray}} \def\ea{\end{eqnarray}} \def\be{\begin{equation}}
\def\ee{\end{equation}} 
\def\gtorder{\mathrel{\raise.3ex\hbox{$>$}\mkern-14mu
\lower0.6ex\hbox{$\sim$}}}
\def\ltorder{\mathrel{\raise.3ex\hbox{$<$}\mkern-14mu
\lower0.6ex\hbox{$\sim$}}}

\usepackage{amssymb,amsbsy,amsmath,amsfonts,amssymb,amscd}
\usepackage{setspace}
\newcommand{\Planck}{\textit{\negthinspace Planck\/}} \newcommand{\planck}{\textit{\negthinspace Planck\/}}
\newcommand{\Spitzer}{\negthinspace \textit{Spitzer\/}}
\newcommand{\Herschel}{\negthinspace \textit{Herschel\/}}
\newcommand{\mission}{{\it PRISM}}
\newcommand{\Euclid}{\negthinspace \textit{Euclid\/}}

\def\m{\ifmmode $m$\else \,m\fi}

\def\st{\ifmmode ^{\mathrm{st}} \else $^{\mathrm{st}}$\fi}
\def\nd{\ifmmode ^{\mathrm{nd}} \else $^{\mathrm{nd}}$\fi}
\def\rd{\ifmmode ^{\mathrm{rd}} \else $^{\mathrm{rd}}$\fi}
\def\th{\ifmmode ^{\mathrm{th}} \else $^{\mathrm{th}}$\fi}

\newcommand\ltsima{$\; \buildrel < \over \sim \;$}
\newcommand\simlt{\lower.5ex\hbox{\ltsima}}
\newcommand\gtsima{$\; \buildrel > \over \sim \;$}
\newcommand\simgt{\lower.5ex\hbox{\gtsima}}
\newcommand\simprop{\lower.5ex\hbox{$\; \buildrel \propto \over \sim \;$}}
\newcommand\mypar[1]{\smallskip\par\noindent\textbf{#1}}

\newcommand{\Msolar}{M_\odot}

\begin{document}

%\AddToShipoutPicture*{\put(75,-100){\includegraphics[width=220mm,height=292mm]%
\clearpage\thispagestyle{empty}
\AddToShipoutPicture*{\put(78,-100){\includegraphics[trim=0cm 0cm 0cm 0cm, clip=true, width=211mm,height=298mm]%
{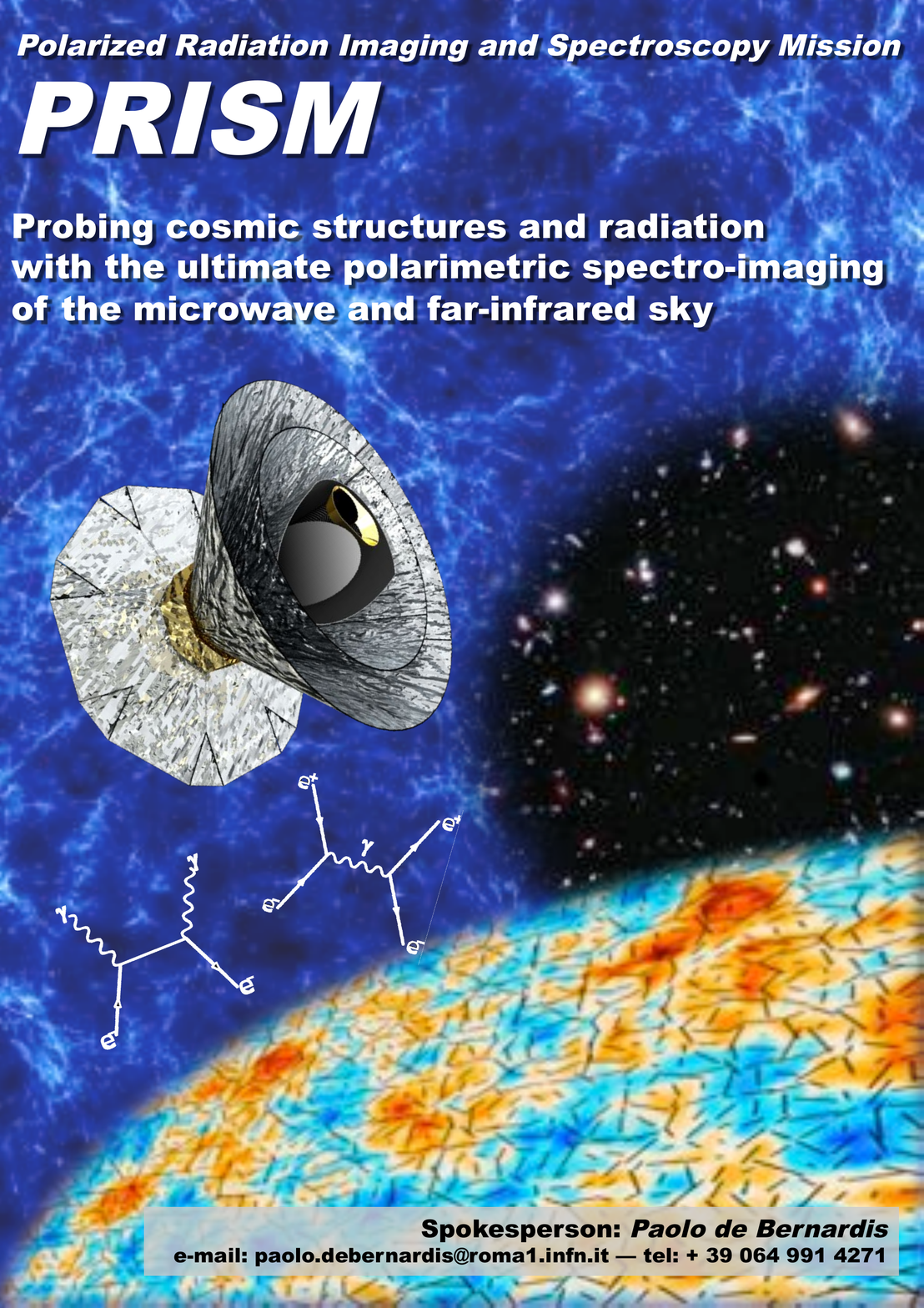}}}%
%{Figures/cover_page.pdf}}}%
\phantom{Invisible, but important}
\newpage 
%\includegraphics[width=0.89\hsize]{Core_cover4.jpg}

%\pagestyle{empty}
%\newpage

\vfill\eject
\pagestyle{plain}

\vfill\eject

\setcounter{tocdepth}{3}
%\begin{spacing}{0.8}
%\tableofcontents
%%\end{spacing}
\clearpage\thispagestyle{empty}
\input authors.tex

\vfill\eject
\setcounter{page}{1}

%====================================================================================
\section{Executive summary}
\input executive_summary.tex

%====================================================================================
\section{Legacy archive}
\input scientific_objectives.tex

%Delabrouille (coord.), Bucher, Chluba, de Zotti, Ferreira, Challinor, Boulanger, ... 

%====================================================================================
\section{Probing the Universe with galaxy clusters}
\input clusters.tex

%Bartlett (coord), Battye, Delabrouille, Diego, Lesgourgues, Melin, Mohr, Rubino-Martin, Macias, ...

%====================================================================================
\section{Extragalactic sources and the cosmic infrared background }
\label{sec:extragalactic}
\input pointsources.tex

%De Zotti (coord), Bethermin, Dole, Dore, Lagache, Massardi, ...

%====================================================================================
\section{Inflation and CMB primordial B-modes}
\label{sec:CMB-B-modes}
\input prim_b_modes.tex

%Ferreira (coord), Bucher, Dunkley, Galli, Jaffe, Melchiorri, Natoli, Tristram, Verde, Garcia-Bellido, Polenta, ...

%====================================================================================
\section{CMB at high resolution} 
\label{sec:CMB-highres}
\input high_res_plus_lensing.tex

%Challinor (coord), Lesgourgues, Stompor, Wandelt, ...

%====================================================================================
\section{CMB spectral distortions}
\label{sec:CMB-spectrum}
\input spectroscopy.tex

%Chluba, Burigana, Hernandez-Monteagudo, Khatri, Rubino-Martin, Trac, Trombetti, ....

%====================================================================================
\section{Structure of the dusty magnetized Galactic ISM}
\input our_galaxy.tex

\input zodi.tex
\section{Strawman mission concept}
\label{sec:mission}
\input mission_concept.tex

%Maffei (coord.), De Bernardis, Piat, Withington, Tartari, Ghribi, Kuzmin, Ghassin, Pajot, Piccirillo, Bersanelli, Menella, ...

%====================================================================================
\section{Competition and complementarity with other observations} 
\vspace{-2mm}

\input relation_to_other_initiatives.tex

%%%%%%%%%%%%%%%%%%%%%%%%%%%%%%%%%
%\bibliographystyle{aa_arxiv.bst}
%\bibliographystyle{aa}
\setlength{\bibsep}{0.0pt}
\bibliographystyle{modplainnat}

\small
\twocolumn
%\begin{spacing}{0.9}
\bibliography{bibliography,Bprim_refs,high_res_plus_lensing,zodi,ref_point_sources,macias,bexpts,non_gaussianity,clusters,galaxy,Bib_add}
%\end{spacing}

\end{document}

%% file: authors.tex
\noindent
{\Large \bf Authors and contributors}
\vspace{3mm}

\noindent
Philippe Andr\'e,
Carlo Baccigalupi, Domingos Barbosa, James Bartlett, Nicola Bartolo, Elia Battistelli, 
Richard Battye, George Bendo, Jean-Philippe Bernard, Marco Bersanelli, Matthieu B\'ethermin, 
Pawel Bielewicz, Anna Bonaldi, Fran\c cois Bouchet, Fran\c cois Boulanger, Jan Brand, Martin Bucher, Carlo Burigana, 
Zhen-Yi Cai, Viviana Casasola, Guillaume Castex, Anthony Challinor, Jens Chluba, Sergio Colafrancesco, Francesco Cuttaia, 
Giuseppe D'Alessandro, Richard Davis, Miguel de Avillez, Paolo de~Bernardis, Marco de~Petris, Adriano de~Rosa, 
Gianfranco de~Zotti, Jacques Delabrouille, Clive Dickinson, Jose Maria Diego,
Edith Falgarone, Pedro Ferreira, Katia Ferri\`ere, Fabio Finelli, Andrew Fletcher, Gary Fuller, 
Silvia Galli, Ken Ganga, Juan Garc\'ia-Bellido, Adnan Ghribi, Joaquin Gonzalez-Nuevo, Keith Grainge, Alessandro Gruppuso, 
Alex Hall, Carlos Hernandez-Monteagudo, 
Mark Jackson,
Andrew Jaffe, 
Rishi Khatri,
Luca Lamagna, Massimiliano Lattanzi, Paddy Leahy, Michele Liguori, Elisabetta Liuzzo, Marcos Lopez-Caniego,
Juan Macias-Perez, Bruno Maffei, Davide Maino, Silvia Masi, 
Anna Mangilli,
Marcella Massardi, Sabino Matarrese, Alessandro Melchiorri, 
Jean-Baptiste Melin, Aniello Mennella, Arturo Mignano, Marc-Antoine Miville-Desch\^enes,
Federico Nati, Paolo Natoli, Mattia Negrello, Fabio Noviello,
Francesco Paci, Rosita Paladino, Daniela Paoletti, Francesca Perrotta, Francesco Piacentini, Michel Piat, 
Lucio Piccirillo, Giampaolo Pisano, Gianluca Polenta,
Sara Ricciardi, Matthieu Roman, Jose-Alberto Rubino-Martin,
Maria Salatino, Alessandro Schillaci, Paul Shellard, Joseph Silk, Radek Stompor, Rashid Sunyaev,
Andrea Tartari, Luca Terenzi, Luigi Toffolatti, Maurizio Tomasi, Tiziana Trombetti, Marco Tucci,
Bartjan Van~Tent, Licia Verde, 
Ben Wandelt, Stafford Withington

%%%%%%%%%%%%%%%%%%%%%%%%%%%%%%%%%%%%%%%
\vspace{6mm}
\noindent
{\Large \bf Coordination group}
\vspace{3mm}

\noindent
{\it The preparation of this science case submitted to ESA has been coordinated by:}
\vspace{1mm}

\noindent
James Bartlett, Fran\c cois Bouchet, Fran\c cois  Boulanger, Martin Bucher, Anthony Challinor, Jens Chluba, 
Paolo de Bernardis (spokesperson), Gianfranco de Zotti, Jacques Delabrouille (coordinator), Pedro Ferreira, Bruno Maffei

%%%%%%%%%%%%%%%%%%%%%%%%%%%%%%%%%%%%%%%
\vspace{6mm}
\noindent
{\Large \bf Supporters}
\vspace{3mm}

\noindent
As a result of 
the highly successful ESA Planck and Herschel missions, Europe has acquired considerable
scientific and technical expertise in the scientifically fruitful and strategic field of microwave and far-infrared
observations and trained 
a new generation of young European astronomers 
in this area.
The science themes outlined 
in this proposal are the logical next step
that will allow ESA to capitalize on these strengths. 

To demonstrate the breadth of support for PRISM, we are in the 
process of assembling a list of supporters that can be found at the following website:
\textcolor{blue}{\large 
\begin{center}
\url{http://www.prism-mission.org}
\end{center}
}
\noindent
Scientists who believe that ESA should pursue 
as part of its programme the science themes presented in this White Paper are strongly encouraged to visit the website and to sign on as 
supporters.

%%%%%%%%%%%%%%%%%%%%%%%%%%%%%%%%%%%%%%%
\vspace{6mm}
\noindent
{\Large \bf PRISM Steering committee}
\vspace{3mm}

\noindent
{\bf France:} Fran\c cois Bouchet, Martin Bucher, Jacques Delabrouille, Martin Giard

\noindent
{\bf Germany:} Jens Chluba, Rashid Sunyaev

\noindent
{\bf Ireland:} Anthony Murphy

\noindent
{\bf Italy:} Marco Bersanelli, Carlo Burigana, Paolo de Bernardis

\noindent
{\bf Netherlands:} Rien van de Weijgaert

\noindent
{\bf Portugal:} Carlos Martins

\noindent
{\bf Spain:} Enrique Mart\'inez-Gonz\'alez, Jos\'e Alberto Rubi\~no-Mart\'in, Licia Verde

\noindent
{\bf Switzerland:} Martin Kunz

\noindent
{\bf United Kingdom:} Anthony Challinor, Joanna Dunkley, Bruno Maffei

\vfill

\hrule 

\vskip 4pt

\noindent 
This is a corrected version (10 June 2013) of the original document submitted on 24 May 2013 to ESA in 
response to {\sl Call for White Papers for the definition of the L2 and L3 Missions in the ESA Science Programme}
({\color{blue}\url{http://sci.esa.int/Call-WP-L2L3}})

%% file: executive_summary.tex
\mission\ is a large-class mission that will carry out the ultimate survey of the 
microwave to far-infrared sky in both intensity and polarization as well as measure its absolute 
spectrum.  \mission\ will consist of two instruments: (1) a high angular 
resolution polarimetric imager with a 3.5$\,$m telescope cooled to around 4K to 
reduce thermal noise, particularly in the far-infrared bands; and (2) a low 
angular resolution spectrometer to compare the sky frequency spectrum to a nearly 
perfect reference blackbody. 
%The joint exploitation of the data from these co-observing high-performance instruments
The two instruments working in tandem 
will enable \mission\ to make breakthrough 
contributions by answering key questions in many diverse areas of astrophysics 
and fundamental science. A few highlights of the new science with \mission\ 
include:

\mypar{(A) The ultimate galaxy cluster survey:} The 
Sunyaev-Zeldovich (SZ) effect is the method of choice for assembling a catalog of 
clusters at high redshift, of particular interest for cosmology because of the tight correlation between 
integrated $y$-distortion and cluster mass. When \mission\ flies, all-sky cluster
samples (e.g., from {\it eROSITA,} \Euclid) will likely count some $10^5$ 
objects, mostly at $z\!<\!1.$ \mission\ will find 10 times more 
clusters extending to deeper redshifts, with many thousands beyond $z=2.$ In 
fact, \mission\ will detect all clusters in the universe of mass larger than 
$10^{14}\Msolar$, and a large fraction of those with mass above $5\times 
10^{13}\Msolar$. Owing to its exquisite spectral coverage, angular resolution and sensitivity, 
\mission\ will measure the peculiar velocity of hundreds of thousands of clusters using the kinetic SZ effect, 
initiating a new research area: the complete mapping of the large-scale velocity field 
throughout our Hubble volume. In addition, 
\mission\ will also be able to probe the relativistic corrections to the classic 
SZ spectral distortion spectrum, thus measuring the gas temperature. This cluster 
sample will allow us to probe dark energy and better understand structure formation at large 
redshift.

\mypar{(B) Understanding the Cosmic Infrared Background:} 
Most star formation in the universe took place at high redshift. Hidden from 
optical observations by shrouds of dust in distant galaxies, it is visible only in the far infrared 
or in X-rays. Emission from these dusty galaxies constitutes the cosmic 
infrared background (CIB) which \mission, owing to its high sensitivity and angular 
resolution in the far infrared, is uniquely situated to investigate. 
The survey will sharpen and extend to 
higher redshifts the determination of the bolometric luminosity function and 
the clustering properties of star-forming galaxies. Tens of thousands of easily 
recognizable, bright, strongly lensed galaxies and hundreds of the very rare 
maximum starburst galaxies, up to $z>6$, will be detected, providing 
unique information on the history of star formation, the physics of the 
interstellar medium in a variety of conditions up to the most extreme, and the 
growth of large scale structure, including proto-clusters of star-forming 
galaxies. The survey will also probe the evolution 
of radio sources at (sub-)mm wavelengths and provide measurements 
of the spectral energy distribution (SED) of many thousands of radio sources over 
a poorly explored, but crucial frequency range.

\mypar{(C) Detecting inflationary gravity waves:} Present precision measurements of cosmic microwave background 
(CMB) temperature anisotropies lend considerable support to simple models of inflation.
However the  most spectacular prediction of inflation---the generation of gravitational waves 
with wavelengths as large as our present horizon---remains unconfirmed. Several 
initiatives from the ground and from stratospheric  balloons are currently 
underway to attempt to detect these gravitational waves through the B-mode 
spectrum of the CMB polarization. However, they suffer from severe 
handicaps such as limited frequency coverage due to atmospheric opacity, unstable seeing 
conditions, and far sidelobes from the ground. It is only from space that one may hope to 
detect the very low-$\ell $ B-modes due to the re-ionization 
bump. Because of its broad frequency coverage and extreme stability, \mission\ will 
be able to detect B-modes at $5\sigma $ for $r=5\times 10^{-4},$ even under pessimistic 
assumptions concerning the complexity of the  astrophysical foreground emissions that must be reliably removed. 
Moreover, \mission\ will be able to separate and filter out the majority of the lensing signal due to gravitational deflections. 

\mypar{(D) Probe new physics through CMB spectral distortions:} 
The excellent agreement between the microwave sky emission and a perfect 
blackbody observed by the {\it COBE FIRAS} instrument is rightfully highlighted as a 
crucial confirmation of Big Bang cosmology. However theory predicts that at 
higher sensitivity this agreement breaks down. Some of the predicted deviations 
are nearly sure bets. Others provide powerful probes of possible new physics. 
The \mission\ absolute spectrometer will measure the spectrum more than three 
orders of magnitude better than {\it FIRAS.} $y$-distortions from the re-ionized gas as 
well as from hot clusters constitute a certain detection. However 
$\mu$-distortions and more general spectral distortions have the potential to 
uncover decaying dark matter and to probe the primordial power spectrum on very 
small scales that cannot be measured by other means, being contaminated by the 
nonlinearity of gravitational clustering at late times.

\mypar{(E) Probe Galactic astrophysics:} \mission\ will have a major impact on Galactic astrophysics
by  providing a unique set of all-sky maps. 
\mission\  will extend Herschel dust observations to the whole sky and 
will map emission lines key to quantifying physical processes. 
The survey will have the  sensitivity and angular resolution required 
to map dust polarization  down to sub-arcminute resolution even at the Galactic poles. 
No project will provide a comparable perspective on interstellar components over such a wide range of scales. 
The \mission\  data will provide 
unique clues to study the interstellar medium, the Galactic magnetic field, and star formation, and
will address three fundamental questions in Galactic astrophysics: 
What are the processes that structure the interstellar medium?
What role does the magnetic field play in star formation?
What are the processes that determine the composition and evolution of interstellar dust?

\vskip 4pt

\noindent
These are but a few of the highlights of the rich and diverse physics and 
astrophysics that \mission\ will be able to carry out.

%% file: scientific_objectives.tex
The hundreds of intensity and polarization maps of \mission\ %, spanning over two decades in frequency, 
will constitute a legacy archive useful for almost all branches of astronomy for 
decades to come. 
Combining low resolution spectrometer data and high resolution images from the imager, \mission\ 
will deliver a full spectro-polarimetric 
survey of the full sky from 50$\, \mu$m to 1$\,$cm.
The spectral resolution will range from about 0.5$\,$GHz to 15$\,$GHz at 
1.4$^\circ$ angular resolution, and from $\delta \nu/\nu \approx 0.025$ to $0.25$ at the diffraction 
limit of a 3.5$\,$m telescope (from $\sim 6''$ to $17'$).

We will make public full-sky maps of the absolute temperature of the CMB and of its polarization (at a resolution 
of about 2 arcminutes with a sensitivity of order 1$\,\mu$K or better per resolution element), of the 
emission of all galactic components in absolute intensity and polarization (including main spectral lines), 
and several catalogues of various galactic and extragalactic objects, among which a catalogue of about a million 
galaxy clusters and large groups up to redshift $z=3$ or more.

%% file: clusters.tex
\begin{wrapfigure}{rt}{0.45\textwidth}
\vskip -0.8cm
\centerline{\includegraphics[width=\linewidth]{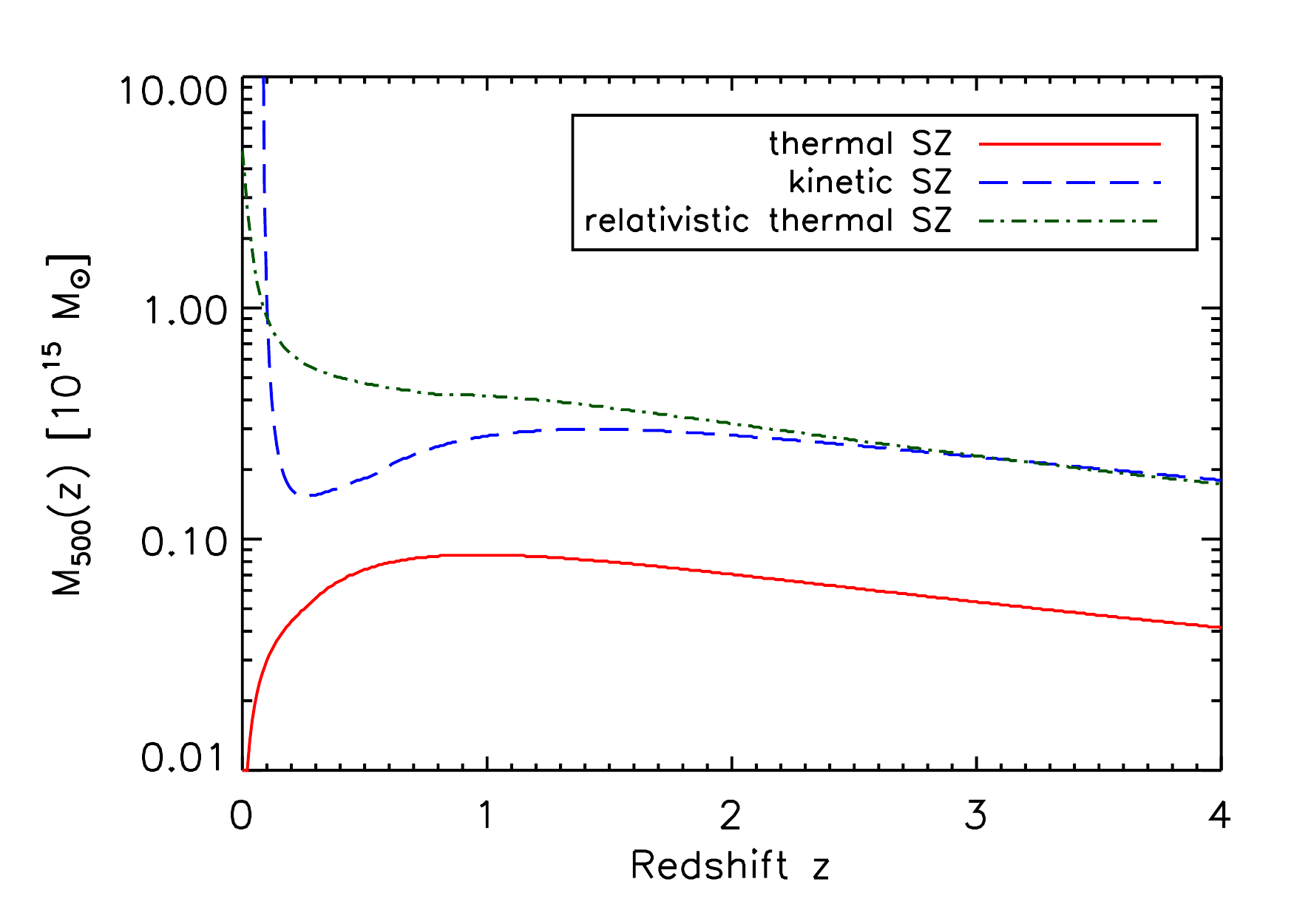}}
\vskip -0.4cm
\caption{{\small Lower mass limits for detection of the indicated SZ effects at signal-to-noise $S/N>5$ as a function of redshift.}}
\label{fig:clusterdetmasses}
\end{wrapfigure}

The \mission\ mission will exploit the advantages of cluster surveying using the SZ effect in a spectacular 
way, surpassing in depth any planned cluster survey and, in addition, achieving an objective unattainable in 
any other way: measurement of the cosmic velocity field throughout the observable universe.  In short, we 
will detect cluster and group systems throughout our Hubble volume from the moment when they first emerge.  
\mission\ will also provide cluster mass determinations out to high redshift through gravitational lensing 
of the CMB in both temperature \citep{seljak2000} and polarization \citep{lewis2006}, something only 
possible because of {\it PRISM}'s high angular resolution and frequency coverage extending into bands 
unreachable from the 
ground. The \Planck, ACT, and SPT experiments demonstrated the potential of the Sunyaev-Zeldovich effect for 
studying galaxy clusters and using them to constrain cosmological models. \mission\ will transform SZ 
cluster studies into arguably our most powerful probe of cosmic large-scale structure and its evolution.

\mypar{The Cluster Catalog and its Applications:} We estimate the content of the \mission\ SZ catalog by 
applying a multi-frequency matched filter \citep{melin2006} to simulations of a typical field at 
intermediate Galactic latitude.  Our detection mass remains below $10^{14}$ solar masses at {\em all 
redshifts} (Fig.~\ref{fig:clusterdetmasses}). Extrapolating from the observed \Planck\ counts, we predict 
nearly $10^{6}$ clusters with many thousands at $z>2$.  We already know from \Planck\ SZ observations 
\citep{planck2011-pepXII,planck2012-pipXI} that the SZ signal in clusters scales as our adopted relation 
down to much smaller masses in the local universe, leaving as our main uncertainty poor knowledge of its 
redshift dependence. This is, of course, the primary motivation for studying the high redshift cluster 
population.

Based on this calculation, \mission\ will surpass all current and planned cluster surveys, including {\it eROSITA}
and Euclid---not just in total numbers, but most importantly in numbers of objects at $z>1.5$.  Cluster 
identification will be vastly more robust for \mission\ than Euclid, which will suffer from the much higher 
contamination rate of optical/NIR cluster searches, especially at redshifts beyond unity.  In all cases, 
only \mission\ has the ability to find significant numbers of clusters in the range $2<z<3$, the critical 
epoch that current observations identify as the emergence of the characteristic cluster galaxy 
population on the red sequence.  {\it PRISM}
will also enable us to explore the abundance of the intra-cluster medium 
(ICM) through the $Y$--$M$ relation and its relation to the galaxy population at these high redshifts.

At the time of operation, large imaging (e.g., {\it DES, LSST, HSC}) and spectroscopic surveys (e.g., {\it 4MOST, PFS, 
WEAVE, BigBOSS/MS-DESI, SKA}) will have covered the entire extragalactic sky.  We will easily be able to 
obtain redshifts, spectroscopic or photometric, for all objects to $z=2$, and the two micron cutoff of 
Euclid's IR photometric survey (H band) will be sufficient to detect the $4000\AA $ break in brighter cluster 
galaxies at higher redshifts.

\mypar{Catalog as a cosmological probe:} As an example of the cosmology constraints that can be 
obtained from the expected cluster catalog, we performed a standard Fisher analysis to constrain four 
parameters, $\Omega_{\rm m}$, $\sigma_8$ and the dark energy equation-of-state parameters $w_0$ and $w_1$, 
in a standard flat $\Lambda$CDM cosmological model.  The constraints on the latter dark energy parameters 
are $w_0=-1\pm 0.003$ and $w_1=0\pm 0.1$ after marginalization over the first two parameters.  
Despite its simplicity, this example nevertheless illustrates the power of the {\it PRISM} cluster catalog as a 
dark energy probe.

\mypar{Cosmic velocity field:} \mission\ will initiate an untapped research area: study of the velocity 
field through the kinetic SZ effect \citep{sz1972,rephaeli1991,birkinshaw1999}, an independent probe of dark 
matter and large-scale structure evolution.  In Fig.~\ref{fig:clusterdetmasses} we show mass limit to which 
we expect to measure a velocity of $300$\,km s$^{-1}$ to five sigma on individual clusters.  This mass limit 
means that we will obtain velocity measurements for hundreds of thousands of clusters out to the highest 
redshifts.  In addition, by comparing measured velocities to mass concentrations, say from Euclid lensing or 
galaxy surveys, we can test the theory of gravity on cosmic scales and to high redshift.  This science is 
unattainable by any other means.

\mypar{Relativistic and non-thermal effects:} We will determine the temperature of clusters down to a mass 
limit just above $10^{14}$ solar masses by measuring the relativistic corrections to the thermal SZ 
spectrum \citep{rephaeli1995,challinor1998,itoh1998,sazonov1998,birkinshaw1999}.  These same characteristics 
allow us to search for non-thermal signatures in the spectra that could signal the presence of highly 
energetic particles, perhaps dark matter annihilation products, and even study the temperature structure of 
the most massive systems.

\mypar{Diffuse SZ and the cosmic web:} The diffuse, unresolved SZ effect probes a different mass and 
redshift range than observations of individually detected objects.  We will study 
this diffuse effect
through the power 
spectrum and higher order moments of an SZ map of the sky.  \Planck\ recently extracted the first Compton 
parameter ($y$-fluctuation) map \citep{planck2013-XXI}, but the results are limited by foregrounds and 
noise. With many more spectral bands and much better sensitivity and resolution, \mission\ will 
significantly improve the results, making possible attempts to directly map the cosmic web (i.e., its 
filaments) over large scales through its diffuse gas content.

We will explore the gas content of dark matter halos down to very low masses, a research area pioneered by 
\Planck\ by stacking SZ measurements based on known objects to detect the signal down below $10^{13}$ solar masses 
\citep{planck2011-pepXII,planck2012-pipXI}.  The measurement over such a vast range is unique to the SZ 
effect and a highly valuable constraint on the mysteries of the feedback mechanisms at the heart of galaxy 
formation.  \mission\ greatly expands this important science area by pushing to the lowest possible masses 
and by probing gas content as a function of object properties.  Coupled with our lensing measurements, we 
have a new and exceptional tool to study the relation between luminous and dark matter.

\mypar{Polarized SZ effect:} \mission\ will enable searches for the polarized SZ effects, giving access to 
transverse cluster velocities and measurements of the CMB quadrupole at distant locations.

%% file: pointsources.tex
\def\lsim{\,\lower2truept\hbox{${<\atop\hbox{\raise4truept\hbox{$\sim$}}}$}\,}
\def\gsim{\,\lower2truept\hbox{${> \atop\hbox{\raise4truept\hbox{$\sim$}}}$}\,}

\begin{SCfigure}[5]
%\begin{center}
\hspace{-1cm}
\begin{minipage}[c]{7.5cm}
	\includegraphics[trim=0.5cm 0.7cm 0.5cm 0cm, clip=true, width=\textwidth]{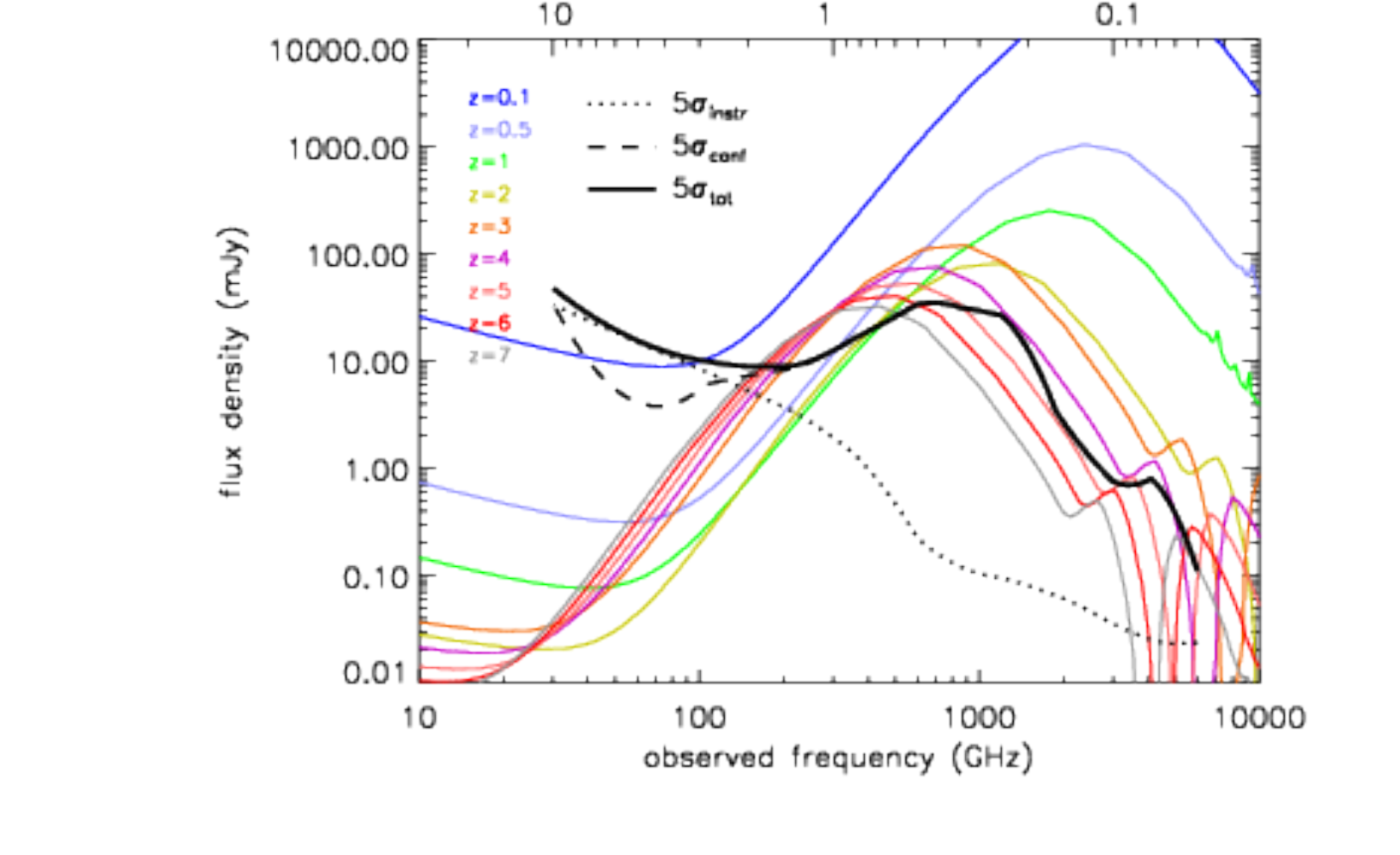}\\
	\includegraphics[trim=0.5cm 0cm 0.5cm 0cm, clip=true, width=\textwidth]{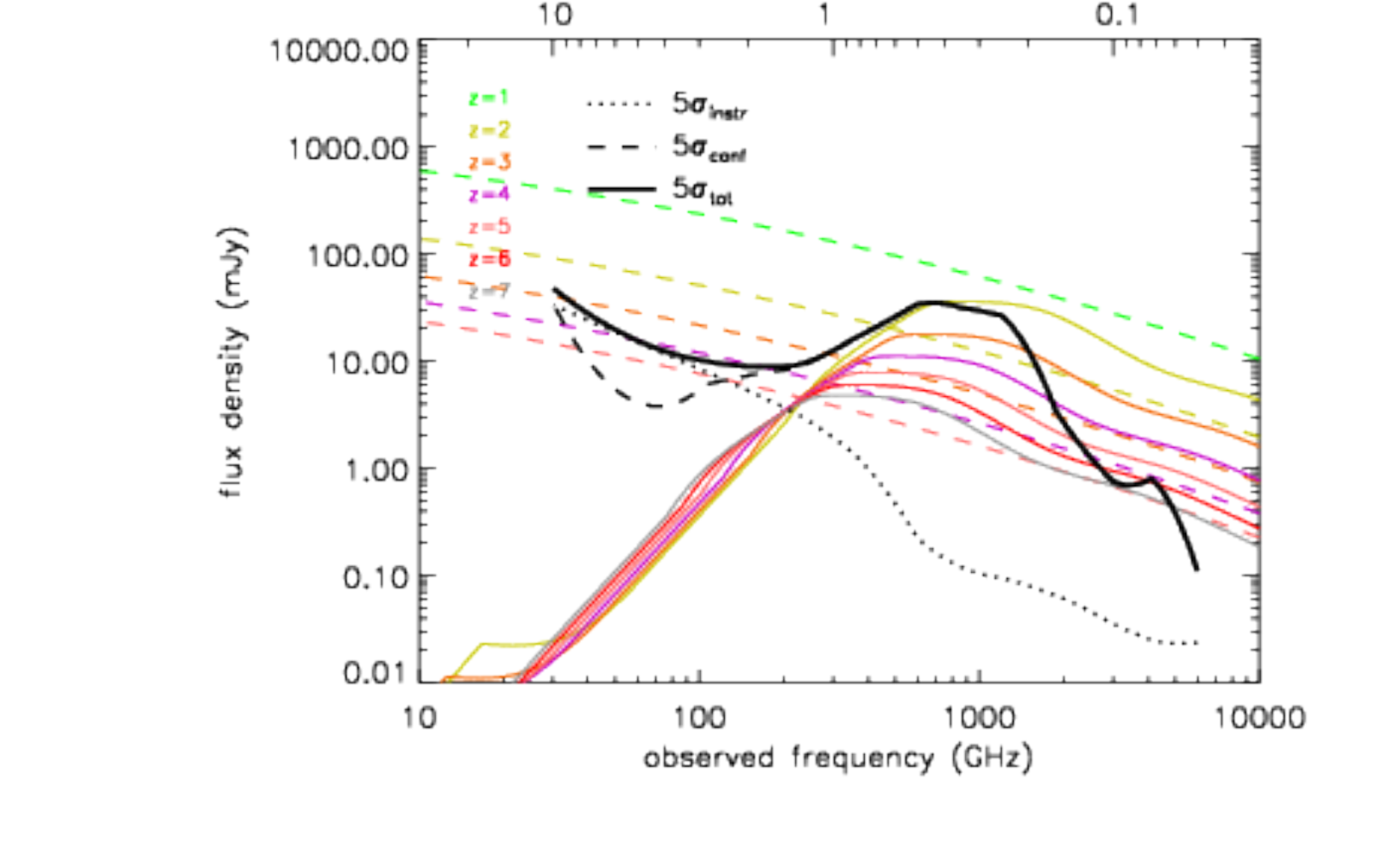}
\end{minipage}
\vspace{-0.9cm}
\caption{\small 
SEDs of dusty galaxies (top panel) and of AGNs (bottom panel) at different redshifts compared with 
estimated $5\sigma$ detection limits (solid black line) taking into account instrumental and 
confusion noise summed in quadrature. The instrumental noise refers to the full mission. The 
$5\sigma$ detection limits allowing for either component are shown by the dotted and the dashed 
black lines, showing that {\mission} is confusion limited above $\approx 150\,$GHz. We have assumed 
that component separation techniques, extensively validated both on simulations and on real data, can 
efficiently remove diffuse emissions such as the CMB (that would otherwise dominate the fluctuation 
field for $\nu\lsim 220\,$GHz) and Galactic emissions. In the top panel, at $z=0.1$ and 0.5 we have 
plotted the Arp\,220 SED scaled to an IR (8--$1000\,\mu$m) luminosity of $10^{12}\,L_\odot$. At $z\ge 
1$ we have used the SED of the $z\approx 2.3$ galaxy SMM~J2135-0102 scaled to $L_{\rm 
IR}=10^{13}\,L_\odot$ for $z=1$ and $z=2$, and to $L_{\rm IR}=3\cdot 10^{13}\,L_\odot$ \citep[the 
luminosity of the $z=6.34$ galaxy detected by \Herschel/SPIRE,][]{Riechers2013} for $z\ge 3$. In the 
bottom panel, the solid colored lines represent SEDs of a type-2 QSO (contribution of the host-galaxy 
subtracted) with $L_{\rm IR}=10^{13}\,L_\odot$ at several redshifts $\ge 2$, while the dashed colored 
lines show a schematic representation of the SED of the prototype blazar 3C\,273 shifted to redshifts 
from 1 to 5. }\label{fig:SED}
\end{SCfigure}

\mypar{Early evolution of galaxies:} Although \Herschel\ and \Spitzer\ made
spectacular advances in 
our understanding of early, dust enshrouded phases of galaxy evolution, our knowledge of 
star-formation history in the distant universe is still very incomplete. The {\mission} mission will 
make essential progress thanks to its unique properties: full sky coverage and unparalleled 
frequency range. As illustrated in Fig.~\ref{fig:SED}, {\it PRISM}'s unprecedented frequency coverage provides 
direct measurements of the bolometric luminosities of star-forming galaxies up to high redshifts. At 
$z\gsim 2$, i.e., in the redshift range where both the cosmic star formation and the accretion rate 
onto supermassive black-holes are maximum, both the IR peak associated with the dusty torus around AGN 
($\lambda_{\rm p,AGN}\approx 30\times(1+z)\,\mu$m) and the peak of dust emission in the host galaxy are 
within the covered range. Moreover, measurements of the complete far-IR to mm-wave SED will 
vastly improve the accuracy of photometric redshift estimates that have a rms error of $\approx  
0.2(1+z)$ with SPIRE alone \citep{Lapi2011}. {This means that the {\mission} survey will allow us to 
characterize to high statistical precision the evolution with redshift of the bolometric luminosity 
function.}  At $z\gsim 2$ it will be possible to investigate the relationships between star-formation 
and nuclear activity: What fraction of the bolometric energy radiated by star-forming galaxies is 
produced by accretion onto supermassive black holes in active galactic nuclei (AGN)? What are the 
evolution properties of far-IR selected AGNs? What fraction of them is associated with active star 
formation? Are the growth of central super-massive black hole formation and the build-up of stellar 
populations coeval? The substantially higher spatial resolution (thanks to the shorter wavelength 
channels) and the correspondingly higher positional accuracy compared to \Herschel/SPIRE will greatly 
improve identification of reliable counterparts in other wavebands, necessary for a comprehensive 
understanding of the properties of detected galaxies.

Its all-sky coverage makes {\mission} uniquely suited to study rare phenomena. Examples are the 
`maximum starburst' galaxy at $z=6.34$, detected by \Herschel/SPIRE \citep{Riechers2013}, or the most 
luminous star-forming hyper-luminous IR galaxies, such as the binary one, pinpointing a cluster of 
star-bursting proto-ellipticals at $z=2.41$ discovered by \citet{Ivison2013}.  The $z=6.34$ galaxy was 
found when looking for {\it ultra-red} sources with flux densities $S_{250\mu\rm m} < S_{350\mu\rm m} < 
S_{500\mu\rm m}$. The {\mission} survey will allow us to look for even redder sources, potentially at 
even higher redshifts, and will provide a test of our understanding of the interstellar medium and of 
star-formation under extreme conditions.

Strongly gravitationally lensed systems have long been very difficult to identify in sufficiently 
large numbers to be statistically useful. This situation changed drastically with the advent of 
(sub-)mm surveys. One of the most exciting \textit{Herschel}/SPIRE results was the direct 
observational confirmation that almost all the galaxies brighter than $\approx 100\,$mJy at $500\,\mu$m 
are either strongly lensed or easily identifiable low-$z$ spirals \citep{Negrello2010}. The surface 
density of strongly lensed high-$z$ galaxies above this limit is $\approx 0.3\,\hbox{deg}^{-2}$, 
implying that an all-sky survey can detect $\sim 10^4$ such systems. The fact that these sources are 
very bright makes redshift measurements with CO spectrometers and high resolution imaging with 
millimeter interferometers relatively easy. {This will allow us to get detailed information on 
obscured star formation in the early Universe and the on processes driving it in observing times hundreds 
of times shorter than would be possible without the help of gravitational amplification and with an 
effective source-plane resolution several times higher than could otherwise be achieved.}

%%%%%%%%%%%%%%%%%%%%%%%%%%%%%%%%%%%%%%%%%%

Large numbers of strongly lensed galaxies are also expected from large area optical surveys. It should 
be noted, however, that sub-mm selection has important distinctive properties. The selected lensed 
galaxies are very faint in the optical, while most foreground lenses are passive ellipticals, 
essentially invisible at sub-mm wavelengths so that there is no, or little, contamination between 
images of the source and of the lens. This makes possible the detection of lensing events with small 
impact parameters. Also, compared to the optical selection, (sub-)mm selection allows us to probe 
earlier phases of galaxy evolution.

Optical spectroscopy of galaxies acting as lenses can be exploited to measure the mass distribution of 
their dark matter halos as a function of redshift. Note that {\Euclid} will directly provide redshifts 
for the majority of the lenses out to $z\sim 1$ in its area. The large number of 
newly identified strongly lensed galaxies will directly probe the evolution of 
large-scale structure. Large samples of strongly lensed galaxies are also essential for many other 
astrophysical and cosmological applications \citep{Treu2010}.

{\mission} will study the angular correlation function of detected sources with much better statistics 
than was possible with \Herschel 's extragalactic surveys that, altogether, cover little more than 2\% of 
the sky. Also, the accurate photometric redshifts will allow us to follow evolution with cosmic time. 
Clustering properties measure the mass of dark matter halos associated with galaxies and are a 
powerful discriminant for galaxy formation and evolution models. Studies of the correlation function 
of the power spectrum also establish occupation numbers of star-forming galaxies, and therefore 
their environments. In particular this study will allow us to detect high-$z$ {\it proto-clusters} of 
dusty galaxies. We thus investigate an earlier evolutionary phase of the most massive 
virialized structures in the Universe. This science is possible only in the wavebands covered
by {\it PRISM.}

The {\mission} clustering data will extend to much higher redshift than {\Euclid}, 
whose wide-area survey will accurately map the galaxy distribution up to $z\sim 1$. The {\mission} 
data will provide information at higher $z$, and primarily over the redshift range $2<z<3$, 
corresponding to the peak in star formation activity. Moreover, optical and near-IR data severely 
underestimate the SFR of dust obscured starbursts and may entirely miss these objects, which are the 
main targets of far-IR/sub-mm surveys such as {\mission}. {Only the combination of {\mission} and 
{\Euclid} data will provide a complete view of the spatial distribution of galaxies and of how star 
formation is distributed among dark matter halos.}

The {\mission} sensitivity and spectral coverage will allow substantially improved measurements of the 
cosmic infrared background (CIB) spectrum with an accurate removal of all contaminating signals. 
{\mission} will also measure in a uniform way the CIB power spectrum over an unprecedented range of 
frequencies and of angular scales (from $\sim 10\,$arcsec to tens of degrees).

\mypar{Radio sources:} {\mission} will extend the counts of radio sources, both in total and in 
polarized intensity, by at least one order of magnitude downwards in flux density compared to \Planck. 
Above 217 GHz, the counts will be determined for the first time over a substantial flux density range 
with good statistics. This will make possible the first investigation of the evolutionary properties of radio 
sources at (sub-)mm wavelengths. {\mission} will provide measurements of the spectral energy 
distribution (SED) of many thousands of radio sources and of multifrequency polarization properties 
for hundreds of them. The vast majority of these sources are expected to be blazars, and the accurate 
determination of their spectra will allow us to understand how physical processes occurring along 
relativistic jets shape the SED. For steep-spectrum sources we will obtain the distribution of break 
frequencies due to electron aging, allowing an unbiased estimate of the distribution of radio source 
ages. Moreover, these observations will shed light on the relationship between nuclear radio emission 
and star formation activity in the host galaxies.

%% file: prim_b_modes.tex
\def\planck{\textit{Planck}}

At the heart of modern cosmology is a set of initial conditions generated at very early times 
by what is known as {\it cosmic inflation}. During inflation, the Universe undergoes a period 
of ultra-rapid accelerated expansion, typically driven by a fundamental scalar field $\phi$, 
with a potential energy $V(\phi)$ that dominates over its kinetic energy. Quantum 
fluctuations of spacetime and the scalar field are amplified and stretched to cosmological 
scales resulting in a quasi-Gaussian stochastic distribution of density perturbations with 
amplitude $A_S$, and a scale dependence characterized by the {\it scalar spectral index}, 
$n_S\equiv1+d\ln A^2_S(k)/d\ln k$. Theory predicts that $A_S$ and $n_S$ depend on the 
details of $V$ and hence $\phi .$ Furthermore, interactions of $\phi$ with itself and with 
other fields induce cross-correlations between perturbation modes, leading to non-Gaussianity 
which can be detected in higher order statistics (bispectrum, trispectrum). Inflation also 
produces a bath of primordial gravitational waves characterized by an amplitude $A_T$ and 
the tensor spectral index $n_T=d\ln A^2_T(k)/d\ln k$. Remarkably, in the simplest models of 
inflation, the ratio between the tensor and scalar perturbations, $r$, is a direct probe of 
$V$ in the early Universe: $r\equiv 16(A_T/A_S)^2\approx M^2_{Pl}(V'/V)^2$. Present 
observations estimate that $V^{1/4}=3.3\times 10^{16} r^{1/4}$ GeV, so that measuring $r$ 
effectively translates into a measurement of the energy scale of inflation. A measurement of 
$r$, $n_S,$ and $n_T$ can directly probe the physics of the early Universe for which there is 
a very rich phenomenology. Single field inflation models can relate $r$ directly with the 
evolution of $\phi$ at early times. Indeed, for an inflationary expansion lasting long enough to 
provide the observed level of homogeneity and isotropy, we have $\Delta \phi/m_{\rm 
Pl}\simeq (r/0.01)^{1/2}$. Multiple field inflation models arising in string theory and other 
proposals for unification at high energies, as well as particle and string production during 
the inflationary period, can lead to even higher values of $r$.

Primordial gravitational waves imprint a unique, as yet undetected, signature in the CMB 
polarization. CMB polarization is a spin-two field on the sky, and is decomposed into the 
equivalent of a gradient---the E-mode---and a curl---the B-mode. Gravitational wave fluctuations 
are visible as the B-mode polarization of the CMB and are the only primordial contribution to 
B relevant at the time of recombination. Hence a detection of B-modes is a direct probe of 
$r,$ and thus the energy scale of inflation and other primordial energetic processes. 
Furthermore, in the simple case of slow-roll inflation we have that $r\approx-8n_T$. 
Additional detailed measurements of the shape of the temperature and polarization spectra 
will measure higher derivatives of the inflationary potential.

\begin{figure}
\begin{center}
\includegraphics[trim=0cm 0cm 0cm 0cm, clip=true,width=10cm]{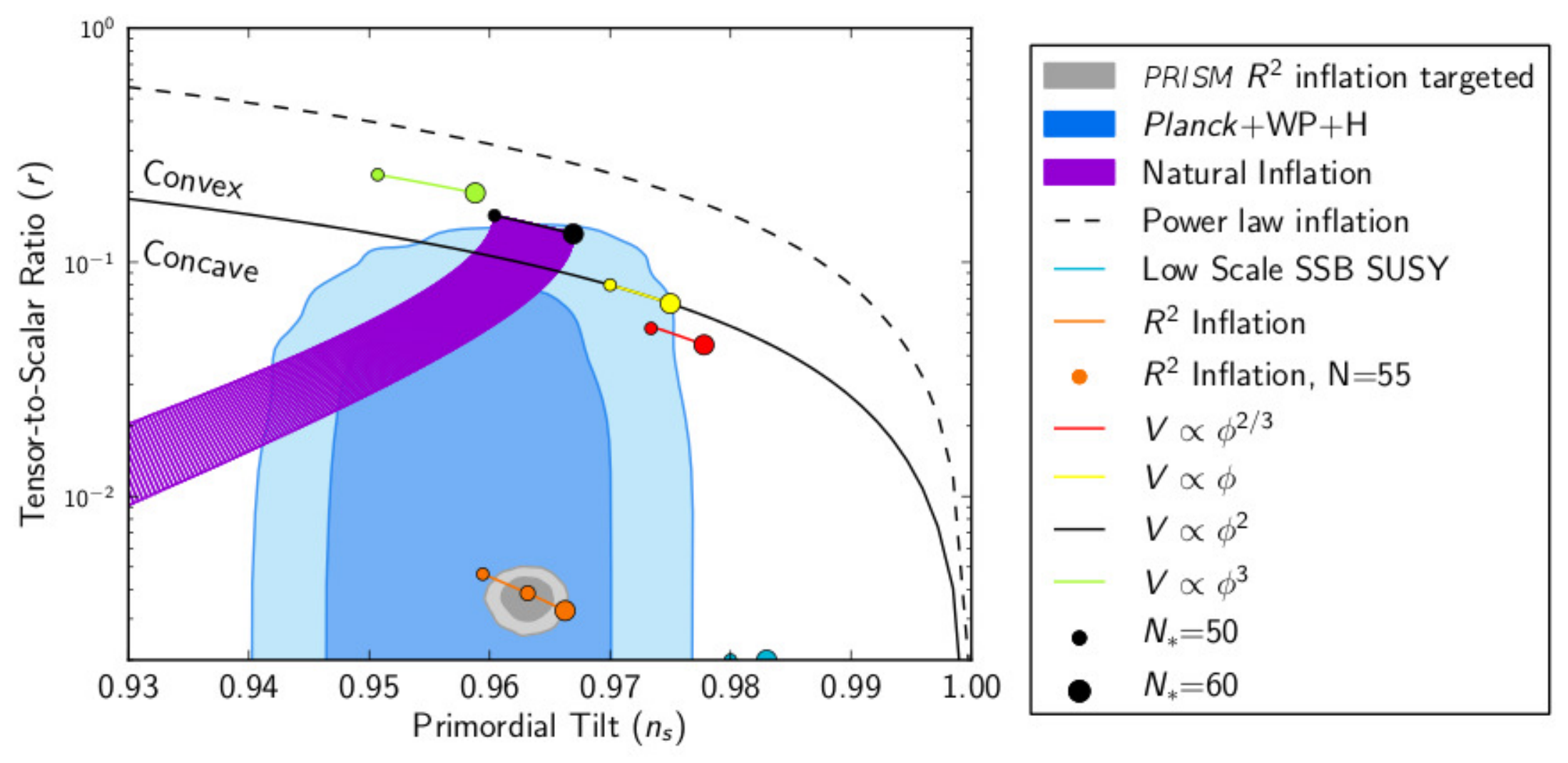}
\hspace{5mm}
\includegraphics[trim=0cm 0cm 5cm 0cm, clip=true,width=6.5cm,height=5cm]{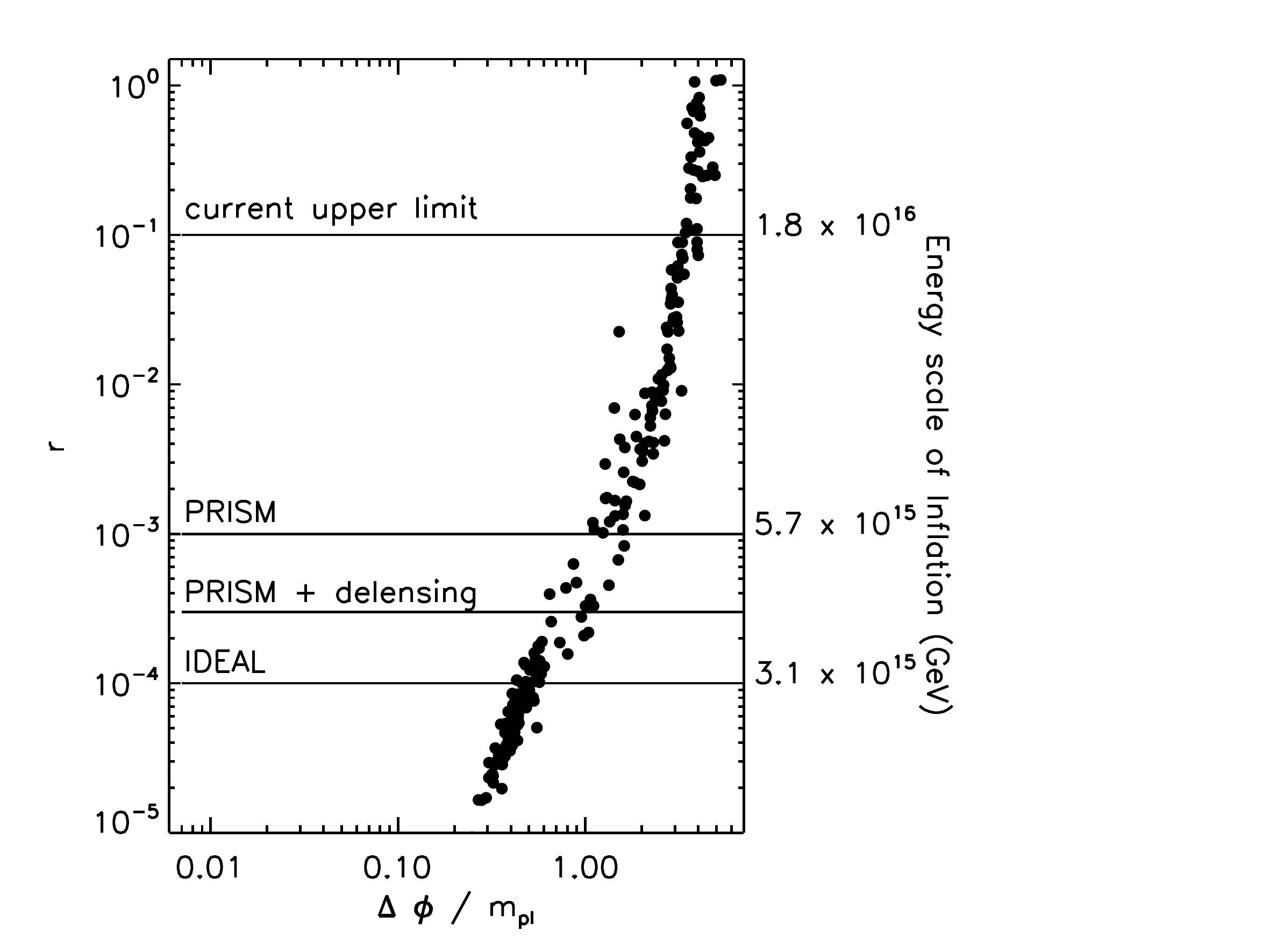}
\caption{\footnotesize Left: Constraints on inflationary potentials from \planck\ and the predicted constraints from {\mission} (not assuming de-lensing) for a fiducial value of $r=5\times 10^{-2}$ (adapted from \citep{planck2013-p17}).  Right: distribution of inflationary model parameters generated using a model independent approach that Monte-Carlo samples the  inflationary flow equations. 
While these simulations cannot be interpreted in a statistical way (e.g., \citet{19, 26, 27}), they show that models cluster around attractor regions (adapted from \citep{ultimatepol}).
}
\label{fig:Bmodes}
\end{center}
\end{figure}

The 2013 \planck\ data release has significantly improved previous constraints on 
inflationary models. In particular, and in the context of the simplest $\Lambda$CDM scenario, 
\planck\ results provide $n_S=0.9624\pm0.0075$ and $r<0.12$. These results are notable 
because
exact scale invariance (i.e., $n_S=1$) of primordial perturbations is ruled out 
at more than $5\,\sigma$. When specific inflationary models are considered, \planck\ imposes 
significant constraints on the potential (Fig.~\ref{fig:Bmodes}), as discussed in 
Ref.~\citep{planck2013-p17}. Indeed \planck\ has shown that it is possible to test many 
inflation models using the CMB temperature data, yet even a forecast \planck\ limit $r < 
0.05$ would leave many interesting models unprobed. Given that the stochastic background of 
gravity waves is the smoking gun of inflation, it is crucial to map as accurately as 
possible the CMB polarization and in particular characterize the B-mode angular power 
spectrum.

%%%%%%%%%%%%%%%%%%%%%%%%%%%%%%%%%%%%%%%%%%%%%%%%%%%%%%%%%%%%%%%%%%%%
%%%%%%%%%%%%%%%%%%%%%%%%%%%%%%%%%%%%%%%%%%%%%%%%%%%%%%%%%%%%%%%%%%%%

To forecast how well we would be able to measure the power spectrum of the B-modes, it is 
important to recognize that the foreground signal is likely to dominate the cosmological 
signal at low $\ell$, where the most constraining information on $r$ is situated.
If we propagate the uncertainties connected to foreground contamination into the parameter error forecasts \citep{ultimatepol,baumanncmbpol,coreWP}, we find that
the proposed experimental set-up will enable us to  explore most large field (single field) inflation models (i.e., where the field moves for $\ge $M$_P$) and to  rule in or out all large-field models, as illustrated in the right-hand panel of Fig.~\ref{fig:Bmodes}. 

As the work by \citet{smith_etal_2008} indicates (see Fig. 8), the instrumental sensitivity, 
angular resolution and, as a result, foreground control and subtraction will enable us to 
achieve a detailed mapping of the lensing signal, and in particular to implement de-lensing 
techniques for the measurement of $r$, improving by a factor of three our constraint on $r$. 
This implies that {\mission} will detect $r\sim 3\times 10^{-4}$ at more than 3$\sigma$. This 
performance is very close, within factors ${\cal O}(1)$, to what an ideal experiment (i.e., 
with no noise and no foregrounds) could achieve, allowing {\mission} to {\it directly} probe 
physics at an energy scale a staggering twelve orders of magnitude higher than the 
center-of-mass energy at the Large Hadron Collider (LHC).

%% file: high_res_plus_lensing.tex
The temperature anisotropies of the CMB have proved to be a remarkably clean probe of the high-redshift universe and have allowed the standard cosmological model to be tested to high precision. However, the accuracy of the recent results from \Planck, based on the temperature anisotropies, are now close to being limited by errors in modelling extragalactic foregrounds. Fortunately, further progress can be made with the polarization anisotropies on small angular scales since the degree of polarization of the anisotropies is relatively larger there (around 4\% by $l=2000$) than the foreground emission. By surveying the full polarized sky in many frequency bands, and with uniform calibration, \mission\ will fully exploit the small-scale polarization of the CMB, improving significantly on results currently obtained from the temperature and those conceivably obtainable in the future with ground-based experiments.

\begin{SCfigure}[1.5]
\includegraphics[width=5.5cm,height=9cm,angle=-90]{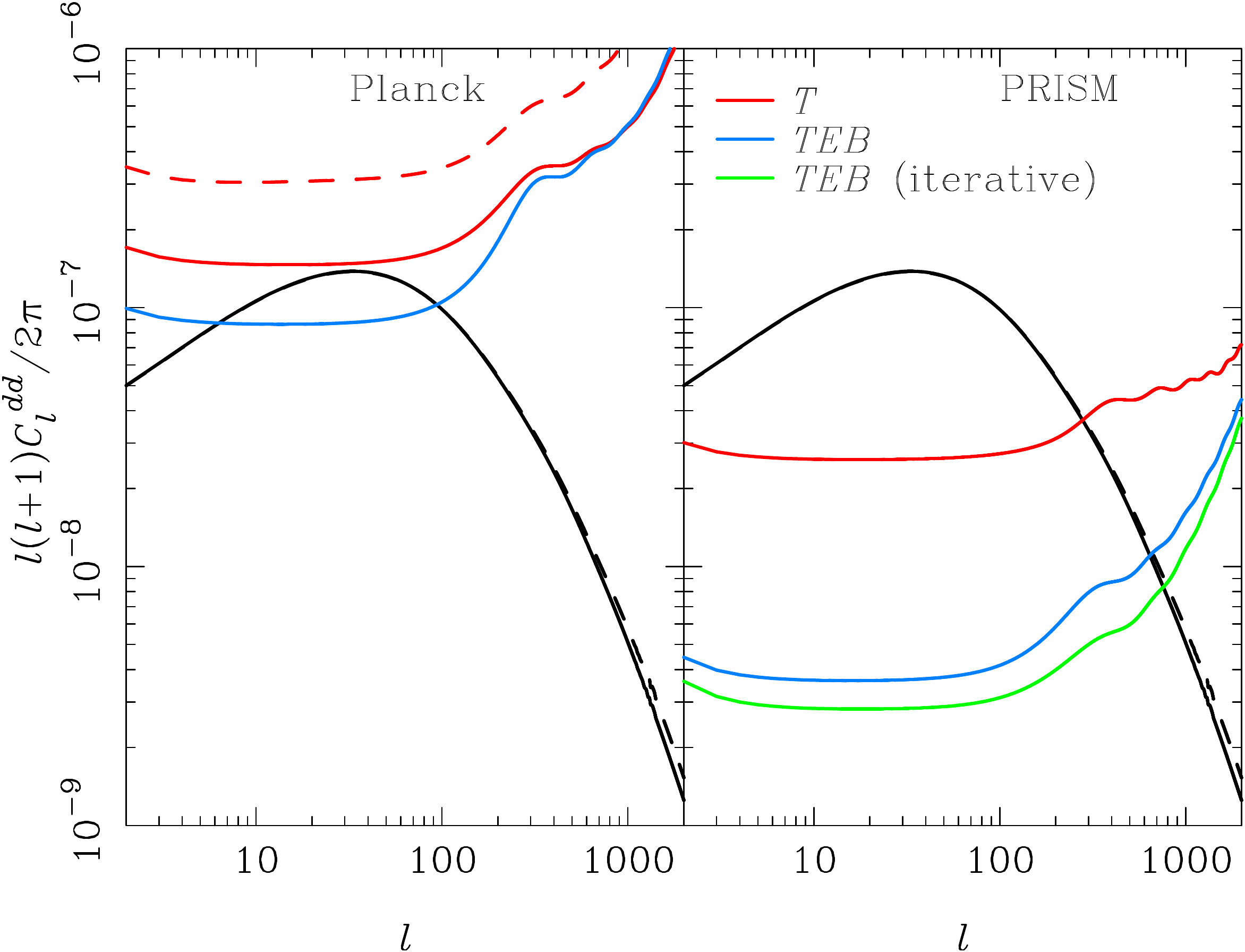}\vspace{-5mm}
\caption{\footnotesize Reconstruction noise on the lensing deflection power spectrum
forecast for the full \Planck\ mission (four surveys; left) and \mission\  (right) using temperature alone (red) and temperature and polarization (blue). For \Planck\ we also show the approximate noise level for the temperature analysis of the nominal-mission data (red dashed)~\cite{2013arXiv1303.5077P}, and for \mission, we also show the approximate noise level (green) for an improved iterative version of the reconstruction estimator.
The deflection power spectrum is plotted based on the linear matter power spectrum (black
solid) and with non-linear corrections (black dashed).
\\
}
\label{fig:CMBlensing}
\end{SCfigure}

\mypar{Probing the dark universe with CMB lensing:}
Gravitational lensing of the CMB provides a clean probe of matter clustering integrated to high redshift.
Lensing can be reconstructed from the CMB anisotropies via specific non-Gaussian 
signatures imprinted by the lenses. \Planck\ has detected lensing via this route at the $25\sigma$ level using the temperature anisotropies, but with low $S/N$ per lensing mode. Polarization-based reconstructions from \mission\ will be a major advance over \Planck, achieving $S/N\gg 1$ over individual multipoles up to $l \approx 600$ over nearly the full sky (see Fig.~\ref{fig:CMBlensing}). Significantly, \mission\ can extract all of the information in the deflection power spectrum on scales where linear theory is reliable. 
%Recent analyses have further demonstrated the ability to constrain dark parameters, that are geometrically degenerate in the primary CMB fluctuations, from the CMB alone via lensing. For example, \Planck\ determines $\Omega_\Lambda$ at 4\% precision, and $\Omega_K$ to around 1\%, in $\Lambda$CDM models with spatial curvature.
To illustrate the power of the lensing measurements from \mission\ in constraining physics that is inaccessible to the primary anisotropies alone due to degeneracies,
we consider the mass of (light) neutrinos. Oscillation data constrain (squared) mass differences, and provide only lower bounds on the total mass summed over eigenstates: $0.06\,\mathrm{eV}$ and $0.1\,\mathrm{eV}$ for the normal and inverted hierarchy, respectively. These hierarchical limits provide natural targets for absolute mass measurements, but are well below the detection limits of current and future laboratory $\beta$-decay experiments. However, masses of these orders can be probed cosmologically via their effect on the clustering of matter. In $w$CDM models with massive neutrinos, we forecast a 1$\sigma$ error of $0.04\,\mathrm{eV}$ for the summed mass. This constraint can be improved further by combining with near-future BAO measurements, for example by a factor of almost two using BOSS, at which point it becomes possible to distinguish between the normal and inverted hierarchies (in the hierarchical limits)~\cite{Hall:2012kg}. 

Lensing constraints from \mission\ would be highly complementary to those from upcoming optical cosmic shear surveys, e.g., \Euclid. The systematic effects are quite different with non-linearities being much less of an issue for CMB lensing and there are no intrinsic-alignment effects. The combination of the two probes of mass is particularly promising, since it allows calibration of multiplicative bias effects such as due to PSF corrections in the optical. Cross-correlating CMB lensing with other probes of large-scale structure, such as galaxies, the Ly$\alpha$ forest or CIB clustering (see Sec.~\ref{sec:extragalactic}), also has exceptional promise, allowing self-calibration of the tracer's bias relation at the sub-percent level. 

\begin{figure}
\centering
\includegraphics[width=5.0cm]{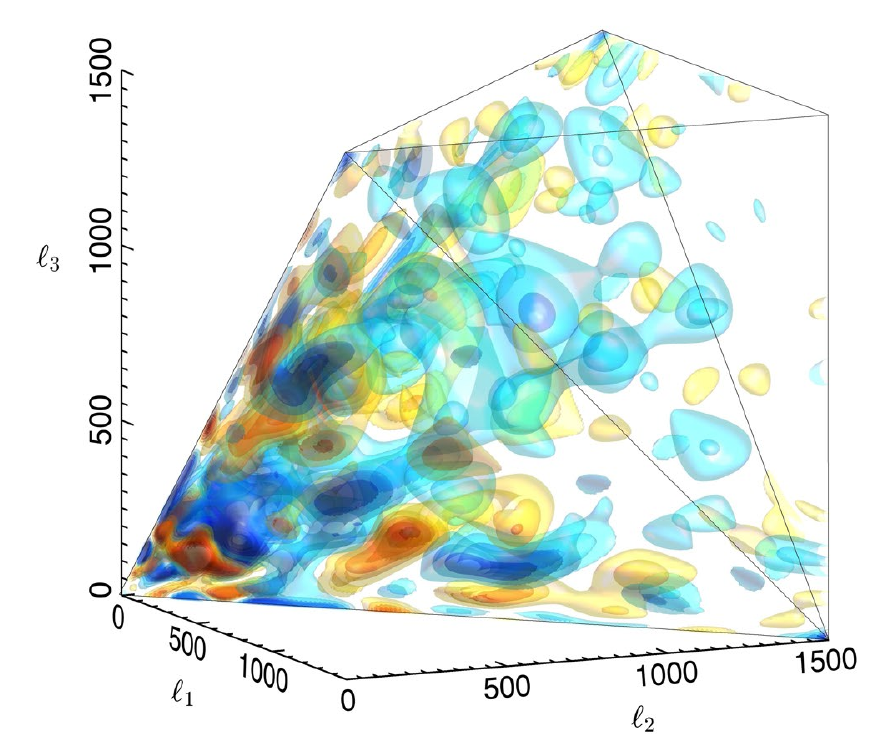}
\includegraphics[width=5.0cm]{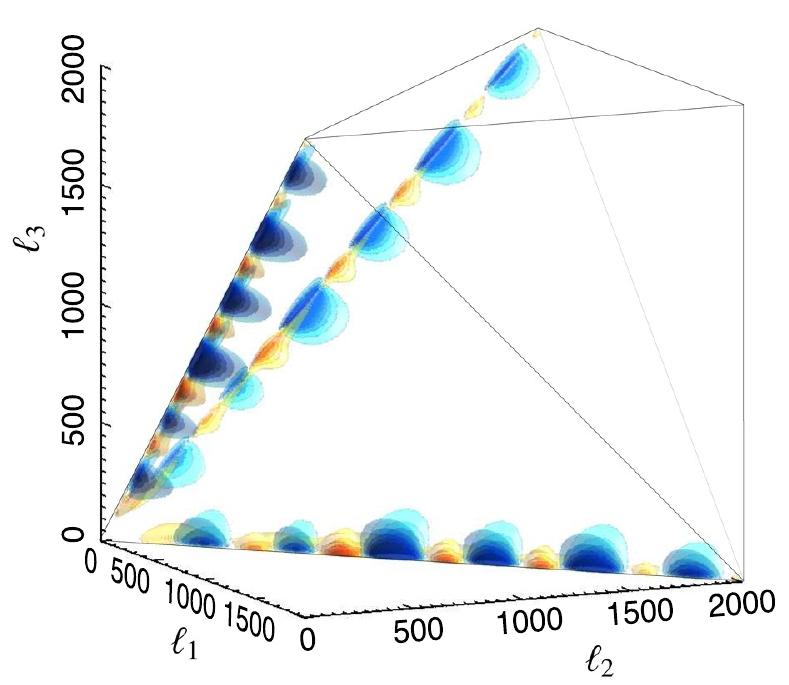}
\includegraphics[width=5.6cm]{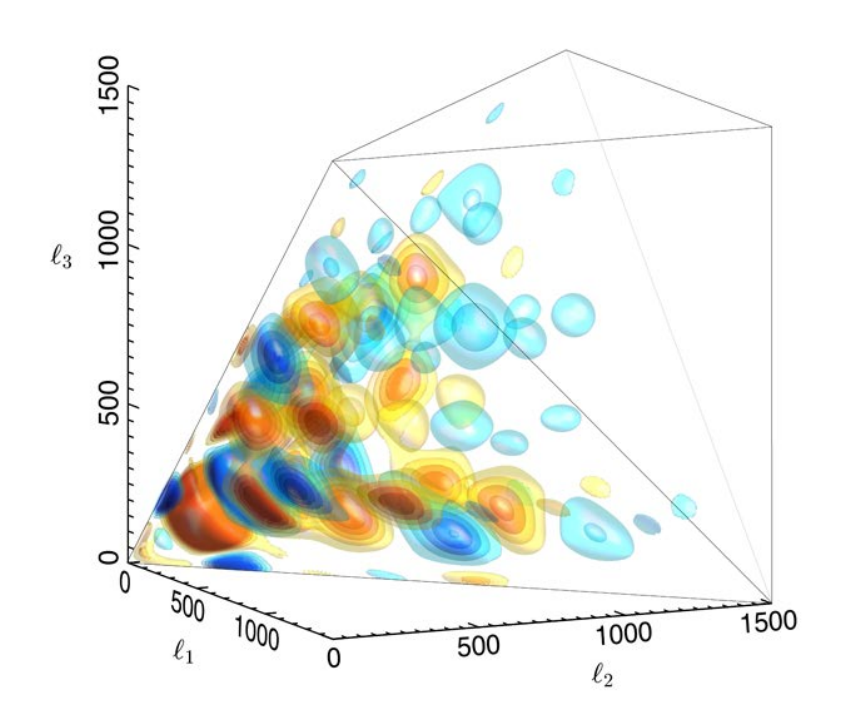}
\caption{\footnotesize \Planck\ CMB temperature bispectrum~\cite{Planck2013ng} (left) and
primordial (right) and late-time (middle) non-Gaussian shapes~\cite{Planck2013ng,2013arXiv1303.5085P}.
%The bispectrum depends on three multipoles $\ell _1, \ell _2, \ell _3$ subject to a triangle constraint, so it contains rich 3D shape information in a tetrahedral domain.   Isocontours are plotted with red positive and blue negative. 
Note the periodic CMB ISW-lensing signal (middle) in the squeezed limit along the edges, which is seen at the 2.5$\sigma$ level in the \Planck\ bispectrum on the left. Scale-invariant signals predicted by many inflationary models are strongly constrained by the \Planck\ bispectrum, although `oscillatory' and `flattened' features hint at new physics. An example of an inflationary `feature' model is shown on the right.
\mission\ will probe these hints with an order of magnitude more resolved triangle configurations.}
%as well as polarization cross-terms.}
\label{fig:recon}
\end{figure}

\mypar{Primordial non-Gaussianity:}
Non-Gaussianity (NG) is now demonstrably a
robust quantitative probe of cosmological physics~\cite{Planck2013ng}.  \Planck\ results dramatically improved previous NG analyses, offering the most stringent test to date of inflationary theory (with $f^{\rm loc}_{\rm NL} = 2.7\pm 5.8$) while also detecting for the first time ISW-lensing and diffuse point source bispectra.   Already \Planck\ offers enticing clues about the nontrivial `shape' of the CMB bispectrum of our universe (see Fig.~\ref{fig:recon}), the origin of which is yet to be explained.   \mission\ would  offer the highest precision reconstructions of the CMB temperature and polarization bispectra and trispectra, which will provide a decisive and unambiguous probe of primordial cosmology back to the Planck era.  At the same time, \mission\ NG data will open new windows for investigating dark energy and gravitational physics, as well as astrophysical sources, large-scale structure and galactic history.

%%%%%%%%%%%%%%%%%%%%%%%%%%%%%%%%%%%%%%%%%%%%%%%%%%%%
%%%%%%%%%%%%%%%%%%%%%%%%%%%%%%%%%%%%%%%%%%%%%%%%%%%%

A unique advantage of the CMB for probing NG is its ability to recognize the distinct 
patterns that physical mechanisms leave in the \textit{shape} of higher-order correlators (Fig.~\ref{fig:recon}). \mission\ will allow a vastly enhanced exploration of physically-predicted NG shapes compared to any 
other projected probe of NG.
%While other probes claim competitive power for detecting `local' bispectral non-Gaussianity, these claims do not yet address the full suite of non-linear systematic and biasing effects.
For example, the
constraint volume in bispectrum space spanned by  the local, equilateral and flattened bispectra will reduce by a factor of 75 compared to the current \Planck\ volume, and a factor of 30 over that predicted from the full-mission \Planck\ data (including polarization). From polarization maps alone (which provide information independent of the 
temperature maps),
we expect a volume reduction factor from the full-mission \Planck\ data to \mission\ of order $110$. Moreover, local-model trispectrum parameters could be measured with a precision $\Delta g_{\rm NL}=3\times 10^4$ and $\Delta \tau_{\rm NL}=1\times10^2$~\cite{smidtetal2010}.
These could investigate consistency conditions between 
polyspectra, which can be used to test large classes of multi-field 
inflation models in addition to single-field inflation.  
There are other alternative inflationary scenarios for which an observable 
non-Gaussian signal is quite natural, e.g., those with features or periodicity in the inflationary potential (Fig.~\ref{fig:recon}). Each of these models has a distinct fingerprint, many uncorrelated with the standard three primordial 
shapes and, in all cases, \mission\ would significantly improve over present \Planck\ constraints, offering genuine discovery potential.   
Beyond searches for primordial NG, \mission\ is guaranteed to make important observations of late-time NG.   For example, it will decisively detect and characterize  the lensing-ISW correlation, driven by dark energy, achieving a $9\sigma$ detection, resulting in a new probe of dark energy physics from the CMB alone.  

\mypar{Parameters from high-resolution polarization spectra:}
\mission\ will measure the CMB angular power spectra with outstanding precision to small angular scales. In particular, in the 105--200\,GHz frequency range, the relatively clean $EE$ polarization spectrum is cosmic-variance limited to $l=2500$ (and the $BB$ spectrum from lensing to $l=1100$).
Such a remarkable measurement of the polarization of the CMB damping tail will be an invaluable source of information on the shape of the primordial power spectrum and the fundamental matter content of the Universe. For example, in $\Lambda$CDM models, the spectral index and its running will be measured more precisely than with current \planck\ data by factors of five and three, respectively. The Hubble constant (a point of tension between \Planck\ data and direct astrophysical measurements) will be measured a factor of 10 better than currently (and $2.5$ times better than expected from the full \planck\ data).
Fundamental questions about the matter content include the effective number of relativistic species $N_{\mathrm{eff}}$, for which a non-standard value (which can relieve the \Planck--$H_0$ tension) could be due to sterile neutrinos, as advocated in particle physics to explain certain anomalies in the neutrino sector, the helium abundance $Y_{\mathrm{P}}$, which provides a clean test of standard BBN, the neutrino mass, and the dark matter annihilation cross-section. 
In one-parameter extensions of $\Lambda$CDM, \mission\ will measure $N_{\mathrm{eff}}$ to 2\% precision and $Y_{\mathrm{P}}$ to 1\%.
These values indicate that a $2\sigma$ anomaly hinted at by \Planck\ could be confirmed decisively with \mission. Moreover, from its measurement of the $B$-mode power spectrum, \mission\ should extend the range of sensitivity to cosmic strings by an order of magnitude over the recent \planck\ constraints~\cite{2013arXiv1303.5085P,Avgoustidis:2011ax}.

%% file: spectroscopy.tex
\newcommand{\zmu}{{z_{\mu}}}
\newcommand{\pot}[2]{#1 \times 10^{#2}}
\newcommand{\ion}[2]{{\text{{\sc #1}\,{\sc #2}}}}
\newcommand{\JC}[1]{\textcolor{blue}{#1}}
\newcommand{\todo}[1]{{\textcolor{red}{[\sc #1]}}}
\newcommand{\Supercore}{{\mission}}

\def\aap{A\&A}
\def\apj{ApJ}
\def\apjs{ApJS}
\def\apjl{ApJL}
\def\mnras{MNRAS}
\def\aj{AJ}
\def\nat{Nature}
\def\aaps{A\&A Supp.}
\def\pra{Phys.Rev.A}         % Physical Review A: General Physics
\def\prb{Phys.Rev.B}         % Physical Review B: Solid State
\def\prc{Phys.Rev.C}         % Physical Review C
\def\prd{Phys.Rev.D}         % Physical Review D
\def\prl{Phys.Rev.Lett}      % Physical Review Letters
\def\araa{ARA\&A}       % Annual Review of Astron and Astrophys
\def\gca{GeCoA}         % Geochimica et Cosmochimica Acta
\def\pasp{PASP}              % Publications of the ASP
\def\pasj{PASJ}              % Publications of the ASJ
\def\apss{Astrophysics and Space Science}
\def\jcap{JCAP}
\def\plb{Phys. Lett. B.}
\def\jhep{JHEP}
% end added by JC
%%%%%%%%%%%%%%%%%%%%%%%%%%%%%%%%%%%%%%%%%%%%%%%%%%%%%%%%%%%%%

%%%%%%%%%%%%%%%%%%%%%%%%%%%%%%%%%%%%%%%%%%%%%%%%%%%%%%%%%%%%%
%\bibliographystyle{plain} % added JC
%\bibliography{Lit}  	    % added JC
%%%%%%%%%%%%%%%%%%%%%%%%%%%%%%%%%%%%%%%%%%%%%%%%%%%%%%%%%%%%%

%[Chluba (coord), Burigana, Hernandez-Monteagudo, Rubino-Martin, Trac, Trombetti, Khatri]
%\vspace{2mm}
%\noindent

\vspace{-1.5mm}
 
{\it COBE}/{\it FIRAS} has shown that the average CMB spectrum is extremely 
close to a perfect blackbody, with possible departures limited to $\Delta I_\nu/I_\nu \lesssim 
\pot{\rm (few)}{-5}$ \citep{Mather1994, Fixsen1996}. 
This places very tight constraints on the thermal history of our 
Universe, ruling out cosmologies with extended periods of significant energy release at 
redshifts $z \lesssim \pot{\rm (few)}{6}$ \citep{Zeldovich1969, Sunyaev1970diss, Illarionov1974, 
Danese1977, Burigana1991, Hu1993, Chluba2005, Chluba2011therm, Khatri2012b}.
There are, however, a large number of astrophysical and cosmological processes that cause 
(inevitable) spectral distortions of the CMB at a level that has only come within reach of 
present-day technology. With \Supercore\ an  
unexplored window to the early universe will be opened, allowing detailed studies of (see 
Fig.~\ref{spectralfigure} for illustration):

%%%%%%%%%%%%%%%%%%%%%%%%%%%%%%%%%%%%%%%%%%
\begin{figure*}
\centering
%\vskip -0.7cm
%\hspace{-3mm}
%\includegraphics[width=0.55\columnwidth]{Figures/monopole_signals.pdf}
%\includegraphics[width=8.2cm, height=7.8cm]{Figures/abun_plot.eps}
%\vspace{-7mm}
%
\hspace{-3mm}
\includegraphics[width=0.5\columnwidth]{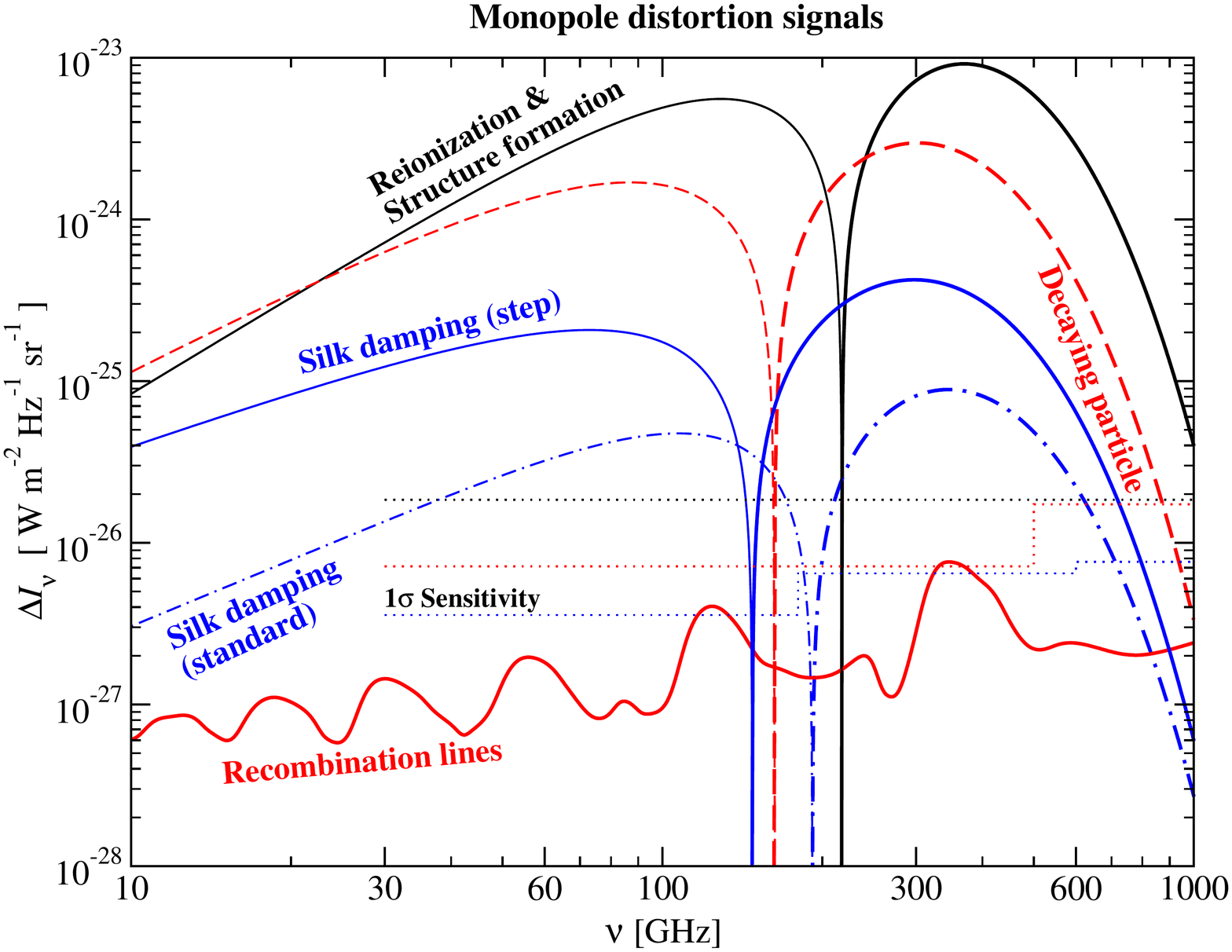}
\includegraphics[width=8.2cm, height=7.2cm]{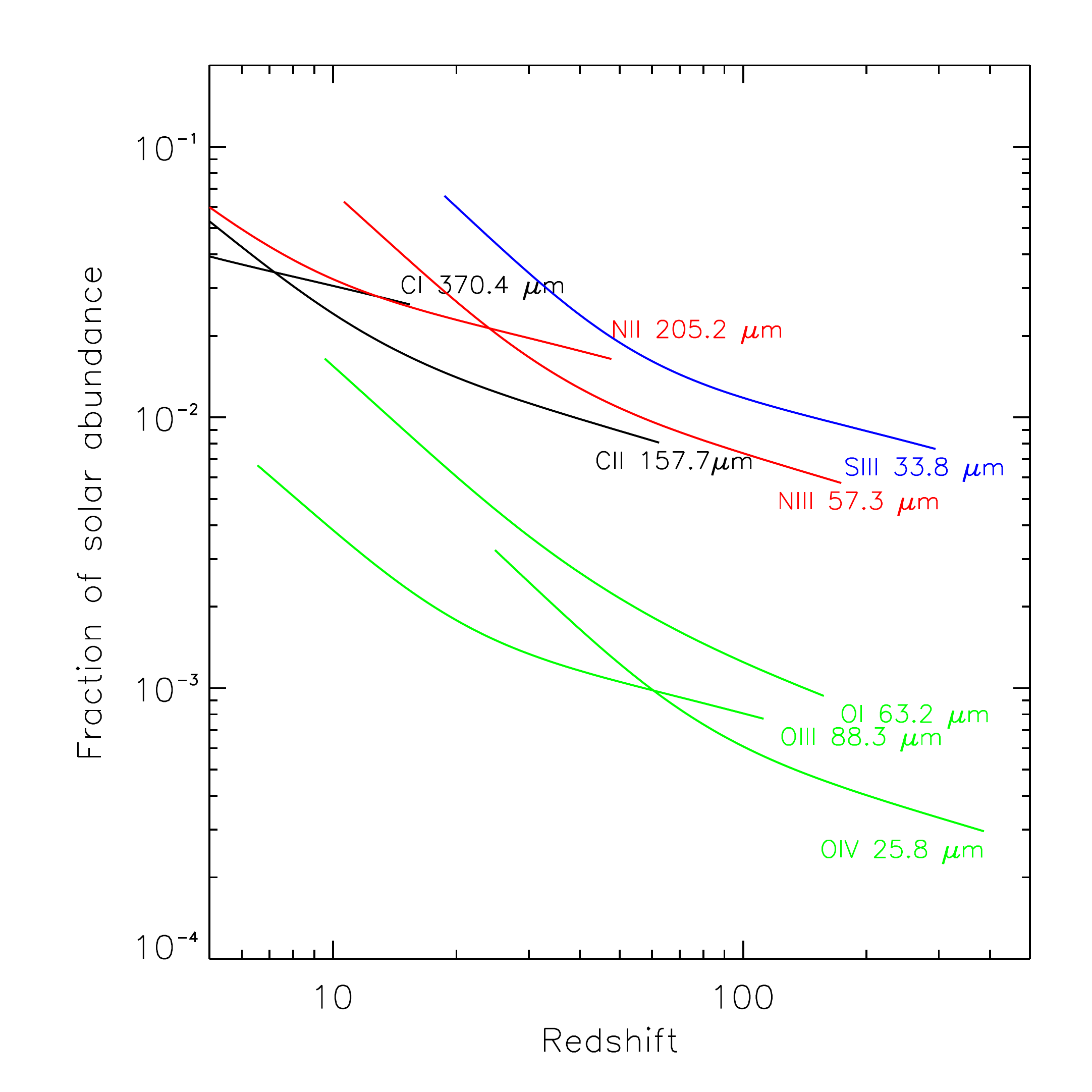}
\vspace{-2mm}
\caption{\footnotesize
Left: spectral distortions for different scenarios. Thick lines 
denote positive, and thinner lines negative signal. The $1\sigma$ sensitivities of 
\Supercore\ for different designs are also indicated. 
Right: projected constraints on different metal ions.}
\label{spectralfigure}
\end{figure*}
%%%%%%%%%%%%%%%%%%%%%%%%%%%%%%%%%%%%%%%%%%

\mypar{Reionization and structure formation:} Radiation from the first stars and galaxies 
\citep{Hu1994pert, Barkana2001}, feedback by supernovae \citep{Oh2003} and structure 
formation shocks \citep{Sunyaev1972b, Cen1999, Miniati2000} heat the IGM at redshifts 
$z\lesssim 10-20$, producing hot electrons that up-scatter CMB photons, giving rise to a 
Compton $y$-distortion with average amplitude $\Delta I_\nu/I_\nu \simeq 10^{-7}-10^{-6}$. 
This signal will be detected at more than a $100\sigma$ by \Supercore, providing a sensitive 
probe of reionization physics and delivering a census of the missing baryons in the local 
Universe. \Supercore\ furthermore has the potential to separate the spatially varying 
signature caused by the WHIM and proto-clusters \citep{Zhang2004}. It also offers a unique 
opportunity to observe the free-free distortion associated with reionization, providing a 
complementary way to study the late evolution of inhomogeneities \citep{ponenteetal2011}.

\mypar{Decaying and annihilating relics:} The CMB spectrum will establish tight limits on 
decaying and annihilating particles in the pre-recombination epoch \citep{Hu1993b, 
daneseburigana94, McDonald2001, Chluba2010a, Chluba2011therm}. This is especially interesting 
for decaying particles with lifetimes $t_{\rm X} \simeq 10^{8}-10^{10}\,{\rm sec}$, as the 
exact shape of the distortion encodes when the decay occurred \citep{Chluba2011therm, 
Khatri2012mix, Chluba2013Green, Chluba2013fore}. \Supercore\ therefore provides an 
unprecedented probe of early-universe particle physics, with many natural particle candidates 
found in supersymmetric models \citep{Feng2003, Feng2010}.

\mypar{Constraining inflation:} Silk damping of small-scale perturbations gives rise to 
CMB distortions \citep{Sunyaev1970diss,Daly1991,Barrow1991,Hu1994} which directly depend on 
the shape and amplitude of the primordial power spectrum at scales $0.6\,{\rm kpc}\lesssim 
\lambda \lesssim 1\,{\rm Mpc}$ (or multipoles $10^5\lesssim \ell \lesssim 10^8$) 
\citep{Chluba2012,Khatri2012short2x2}.
This allows constraining the trajectory of the inflaton at stages unexplored by ongoing or 
planned experiments \citep{Chluba2012inflaton,Powell2012, Khatri2013forecast}, extending our 
reach from 7 $e$-folds of inflation probed with the CMB anisotropies to a total of 17 $e$-folds. 
The signal is also sensitive to the difference between adiabatic and isocurvature 
perturbations \citep{Barrow1991,Hu1994isocurv, Dent2012, Chluba2013iso}, as well as primordial 
non-Gaussianity in the ultra squeezed-limit, leading to a spatially varying spectral signal 
that correlates with CMB temperature anisotropies as large angular scales \citep{Pajer2012, 
Ganc2012}.
A competing monopole signal, characterized by a {\it negative} $\mu$- and $y$-parameter, is introduced by the adiabatic cooling of ordinary matter \citep{Chluba2005, Chluba2011therm, Khatri2011BE}, 
to which \mission\ will also be sensitive.

%%%%%%%%%%%%%%%%%%%%%%%%%%%%%%%%%%%%%%%%%%%
%\begin{wrapfigure}{rt}{0.55\textwidth}
%%\centering
%\vskip -0.4cm
%\centerline{\includegraphics[width=9.5cm,height=5.5cm]{Figures/halpha.pdf}}
%\caption{\footnotesize
%Change in the CMB angular power spectrum (TT) arising from the H$\alpha$
%line generated during recombination, as a function of the redshifted frequency
%(left panel) and the angular multipole (right panel). Solid lines in both
%panels refer to the cases in which the signal is largest. }
%\label{fig:Halpha}
%\end{wrapfigure}
%%%%%%%%%%%%%%%%%%%%%%%%%%%%%%%%%%%%%%%%%%%

\mypar{Metals during the dark ages:} Any scattering of CMB photons after recombination blurs 
CMB anisotropies at small scales, while producing new anisotropies at large scales. Electrons 
from the reionization epoch are the dominant source of optical depth, causing a signature 
already detected by WMAP and Planck 
\citep{WMAP_params, Planck2013params}. The resonant scattering of CMB photon by fine 
structure lines of metals and heavy ions produced by the first stars adds to this optical 
depth, making it frequency-dependent \citep{Basuetal2004}. By comparing CMB temperature and 
polarization anisotropies at different frequencies one can thus determine the abundances of 
ions such as OI, OIII, NII, NIII, CI, CII at different redshifts 
\citep{Hernandezetal2006,Hernandezetal2007a}. Furthermore, UV radiation emitted by the first 
stars can push the OI 63.2\,$\mu$m and CII 157.7\,$\mu$m transitions out of equilibrium with 
the CMB, producing a distortion $\Delta I_\nu/I_\nu \simeq 10^{-8}-10^{-9}$ due to fine 
structure emission \citep{Gongetal2012,Hernandezetal2007b}, providing yet another window to 
reionization within reach of \Supercore.

\mypar{Cosmological recombination radiation:} The recombination of H 
and He 
\citep{Dubrovich1975} at redshifts $z\simeq 10^3-10^4$, corresponding to $\simeq 260\,{\rm kyr}$ 
(\ion{H}{i}), $\simeq 130\,{\rm kyr}$ (\ion{He}{i}), and $\simeq 18\,{\rm kyr}$ 
(\ion{He}{ii}) after the big bang \citep{Jose2006, Chluba2006b, Jose2008}. The 
signal 
provides an independent determination of the cosmological parameters (such as the baryon 
density and {\it pre-stellar} helium abundance) and direct measurements of the recombination 
dynamics, probing the Universe at stages well before the last scattering surface 
\citep{Sunyaev2009}. The effect on the TT power spectrum introduced by resonance scattering 
of CMB photons by the first lines of the Balmer and Paschen series \citep{RHS05, HS05, 
Hernandezetal2007a} will also be detectable with \mission , providing an additional 
opportunity to directly constrain the recombination history and obtain independent 
determinations of cosmological parameters (e.g. $\Omega_{\rm b}$ or $\Omega_{\rm m}$).

\mypar{Non-Gaussianity:} CMB spectral distortions can also provide a new probe of primordial 
NG~\cite{Pajer:2012vz}. We know almost nothing about NG on the small scales that can be 
probed via these observations. In particular, the cross-correlation between $\mu$-type 
distortions and CMB anisotropies is naturally sensitive to the very squeezed limit of the 
primordial bispectrum (probing scales as small as $50 \leq k\,\mathrm{Mpc} \leq 10^4$). Also, 
the power spectrum of $\mu$-distortions can probe the trispectrum of primordial fluctuations. 
Such measurements can be particularly constraining for models where the primordial power 
spectrum grows on small scales~(see e.g.~\cite{Chluba:2012we}), and values $f^{\rm loc}_{\rm 
NL} <1$ can be achieved. Also, $\mu$-type distortions can shed light 
on non-standard initial states for the quantum fluctuations. For a large class of 
inflationary models 
characterized by a non-Bunch-Davies vacuum (whose bispectrum is enhanced in the squeezed 
limit with respect to the local form) a high $S/N$ can be achieved~\cite{Ganc:2012ae}.

\vspace{1mm}\noindent All these examples demonstrate that the CMB spectrum provides a rich 
and unique source of complementary information about the early Universe, with the certainty 
of a detection of spectral distortions at a level within reach of \Supercore's capabilities. 
The CMB spectrum will also establish interesting constraints on the power spectrum of 
small-scale magnetic fields \citep{Jedamzik2000}, cosmic strings \citep{Ostriker1987, 
Tashiro2012, Tashiro2012b}, evaporating primordial black holes \citep{Carr2010}, decay of 
vacuum energy density \citep{BartlettSilk1990, Burigana1993, daneseburigana94}, and other new 
physics \citep{Lochan2012, Bull2013}, to mention a few more exotic examples. Deciphering all 
these signals will be a big challenge for the future. This area has great potential for new 
discoveries and for providing new independent constraints on unexplored processes that cannot be 
explored by other means.

%% file: our_galaxy.tex
%\documentclass[11pt]{article}
%\usepackage{graphicx} 
%%\usepackage{natbib}
%
%\setlength{\topmargin}{-1.9cm}
%\setlength{\oddsidemargin}{0.3cm}
%\setlength{\evensidemargin}{0.3cm}
%\setlength{\textwidth}{15.8cm}
%\setlength{\textheight}{24.5cm}
%\newcommand{\mission}{PRISM}
%
%\begin{document}
%
%\centerline{\Large \bf  The structure of the dusty magnetized Galactic ISM} 
%\bigskip

The data analysis is still on-going but it is already clear that \Herschel\ and \Planck\  will have a profound and lasting impact on  our understanding
of the  interstellar medium and star formation. 
\mission\ holds even greater promise for breakthroughs. 
Dust and synchrotron radiation are the dominant contributions to  the sky emission and polarization to be observed by  \mission . 
Dust emission is an optically thin tracer of the structure of interstellar matter. 
Synchrotron radiation traces the magnetic field over the whole volume of the Galaxy, while 
dust polarization traces the magnetic field within the thin star forming disk, where the interstellar matter is concentrated. 
\mission\ will image these two complementary tracers 
with unprecedented sensitivity and angular resolution. It will also
provide all-sky images of spectral lines, which are key diagnostics of interstellar gas physics. 
No other initiative offers a comparable imaging capability of interstellar components 
over as wide a range of scales. 
In the following subsections we detail 
how \mission\  will address three fundamental questions of Galactic astrophysics: (1) 
What are the processes that structure the interstellar medium? (2)
What role does the magnetic field play in star formation? (3)
What are the processes that determine the composition and evolution of interstellar dust?

\subsection{Structure of interstellar medium} 

\Herschel\ far infrared observations have provided astronomers new insight into 
how turbulence stirs up the interstellar gas, giving rise to a filamentary, web-like structure within the diffuse interstellar medium and  molecular clouds.
\mission\  will extend the \Herschel\ dust observations to the whole sky and provide unique data on emission lines key to quantifying
physical processes.  The spectral range of \mission\  includes
atomic and molecular lines that serve as diagnostics of 
the gas density and temperature, its chemical state, and energy
budget. \Herschel\ has observed these lines along discrete lines of sight with very limited imaging.
By mapping these lines and dust emission over the whole sky at an angular resolution
comparable to that of \Herschel, \mission\  will 
probe the connection between  the structure of matter  and gas cooling across scales. 

The \mission\  sky maps will provide multiple clues to characterize the physical processes that shape
interstellar matter. The CII, CI, and OI fine structure lines and the rotational lines of 
CO and H$_2$O are the main cooling lines of 
cold neutral medium and molecular clouds and probe local physical conditions and the exchange of energy associated with the formation 
of molecular gas within the diffuse interstellar medium and of stars within molecular clouds.   The NII lines at 122 and 205$\, \mu$m are spectroscopic tracers 
of the ionized gas. 
These lines are essential for distinguishing the contribution of neutral and ionized gas to the CII emission. 
\mission\  will have the sensitivity to image the CII line emission at sub-arcminute resolution even at the Galactic poles. 
The CII map can be combined with HI and dust observations
to study the formation of cold gas from the warm neutral medium through the thermal instability.
This analysis will probe the expected link, yet to be confirmed observationally, 
between the small-scale structure of the cold interstellar medium and gas cooling.
The CII line emission is also key to studying the formation of molecular gas by tracing the CO-dark H$_2$ gas \citep{2013arXiv1304.7770P}. 
In star forming molecular clouds, the CO, CI, OI, and H$_2$O lines
are the key tracers of the processes creating  
the initial conditions of star formation and of the feedback from newly formed stars on their parent clouds. 

\subsection{Galactic magnetic field and star formation} 

Star formation results from the action of gravity, 
counteracted by thermal, magnetic, and turbulent pressures \citep{2012A&ARv..20...55H}.  
For stars to form, gravity must locally become the dominant force. 
This happens when the turbulent energy has dissipated and matter has condensed 
without increasing the magnetic field by a comparable amount. 
What are the processes that drive and regulate the rate at which matter reaches this stage?
This is a long standing question to which theorists have over the decades offered multiple explanations, 
focusing on either ambipolar diffusion, turbulence, or 
magnetic reconnection to decouple matter from the magnetic field and allow the formation of 
condensations of gas in which stars may form \citep{2012ARA&A..50...29C}.

\mission\  observations of the
polarization in the far-IR and sub-mm will provide unique clues to understand the role of the magnetic field  
in star formation.  
Compared to  synchrotron radiation and  Faraday rotation, 
%dust polarization in the far-IR and sub-mm is unique for mapping the magnetic field through a tracer of interstellar matter.  
dust polarization images the structure of the magnetic field through an emission process tracing matter. 
It is  best suited to characterize  the 
interplay between turbulence, gravity, and the Galactic magnetic field. 
The \mission\ data will provide unique data to  
study magneto-hydrodynamical turbulence because it will drastically increase the spectral range of accurately probed magneto-hydrodynamical modes. 
The data will provide unprecedented statistical information to characterize  
the energy injection and energy transfer down to the dissipation scales.

Polarization data from the  \mission\  survey will have the  sensitivity and angular resolution required 
to map continuously the Galactic magnetic field  over the whole sky down to sub-arcminute resolution even at the Galactic poles. 
The wide frequency range of the mission 
will measure polarization for separate emission components with distinct temperatures along
the line of sight. \mission\  will provide a new perspective on the structure of the magnetic field 
in molecular clouds, independent of grain alignment,  by 
imaging the polarization of  CO emission in multiple rotational  lines \citep{1981ApJ...243L..75G}.
No  project offers comparable capabilities.  Planck has provided the first all-sky maps of  dust polarization with 5' resolution but the data 
is sensitivity limited even at the highest Planck frequency (353 GHz).  
Ground based telescopes at sub-mm and millimeter wavelengths of bright compact sources at arcsecond resolution (for example with ALMA)
complement the full-sky survey of extended emission from
the diffuse interstellar medium and molecular clouds that only \mission\   can carry out.

\subsection{Nature of interstellar dust}

The combination of spectral and spatial information provided by  \mission\   will provide new tools for studying
the interstellar dust, in particular its nature and its evolution.
Dust properties (e.g., size, temperature, emissivity) are found to vary from one line of sight to another 
within the diffuse interstellar medium and molecular clouds. These observations indicate that dust grains evolve in a manner
depending on their environment  within 
the interstellar medium. They can grow through the formation of refractory or ice mantles,  or 
by coagulation into aggregates in dense and quiescent regions. They can also be destroyed by 
fragmentation and erosion of their mantles under more violent conditions.  The composition of interstellar 
dust reflects the action of interstellar processes, which contribute to breaking and reconstituting grains over timescales 
much shorter than the timescale of injection by  stellar ejecta. While there is broad consensus on 
this view of interstellar dust, the processes that drive its evolution in space are poorly understood \citep{2009ASPC..414..453D}.   
Understanding interstellar dust evolution is a major challenge  in astrophysics
underlying key physical and chemical processes in interstellar space. In particular, to fully exploit the \mission\ data we will need
to characterize where in the interstellar medium grains are aligned with respect to the Galactic magnetic field and with what efficiency. 

Large dust grains (size $> 10$~nm) dominate the dust mass. 
Within the diffuse interstellar medium, these grains are cold ($\sim10-20$~K) and emit within the \mission\ frequency range. 
Dipole emission from small rapidly spinning dust particles constitutes an additional emission component, known as anomalous microwave emission.
Magnetic dipole radiation from thermal fluctuations in magnetic nano-particles
may also be a significant emission component over the frequency range relevant to CMB studies \citep{2013ApJ...765..159D}.
To achieve the \mission\  objectives on CMB polarization, it is  necessary to characterize the spectral dependence of the polarized signal from 
each of these  dust components with high accuracy across the sky. This is a challenge but also a unique opportunity for dust studies.
The spectral energy distribution of dust emission and the polarization signal can be cross-correlated with the spectral diagnostics of 
the interstellar medium structure to characterize the physical processes that determine the composition and evolution of interstellar dust. 
The same data analysis will also elucidate the physics of grain alignment.

%% file: zodi.tex
\mission\ will also probe the zodiacal dust emission from within our solar system.
The fact that PRISM scans a substantial portion of the sky each day allows for a 
three-dimensional tomographic mapping of the zodiacal emission. Understanding zodiacal
emission is crucial both to understanding our solar system and to carrying
out a complete foreground separation.

%% file: mission_concept.tex
\begin{wrapfigure}{rt}{0.44\textwidth}
\vskip -1.4cm
\centerline{\includegraphics[width=\linewidth]{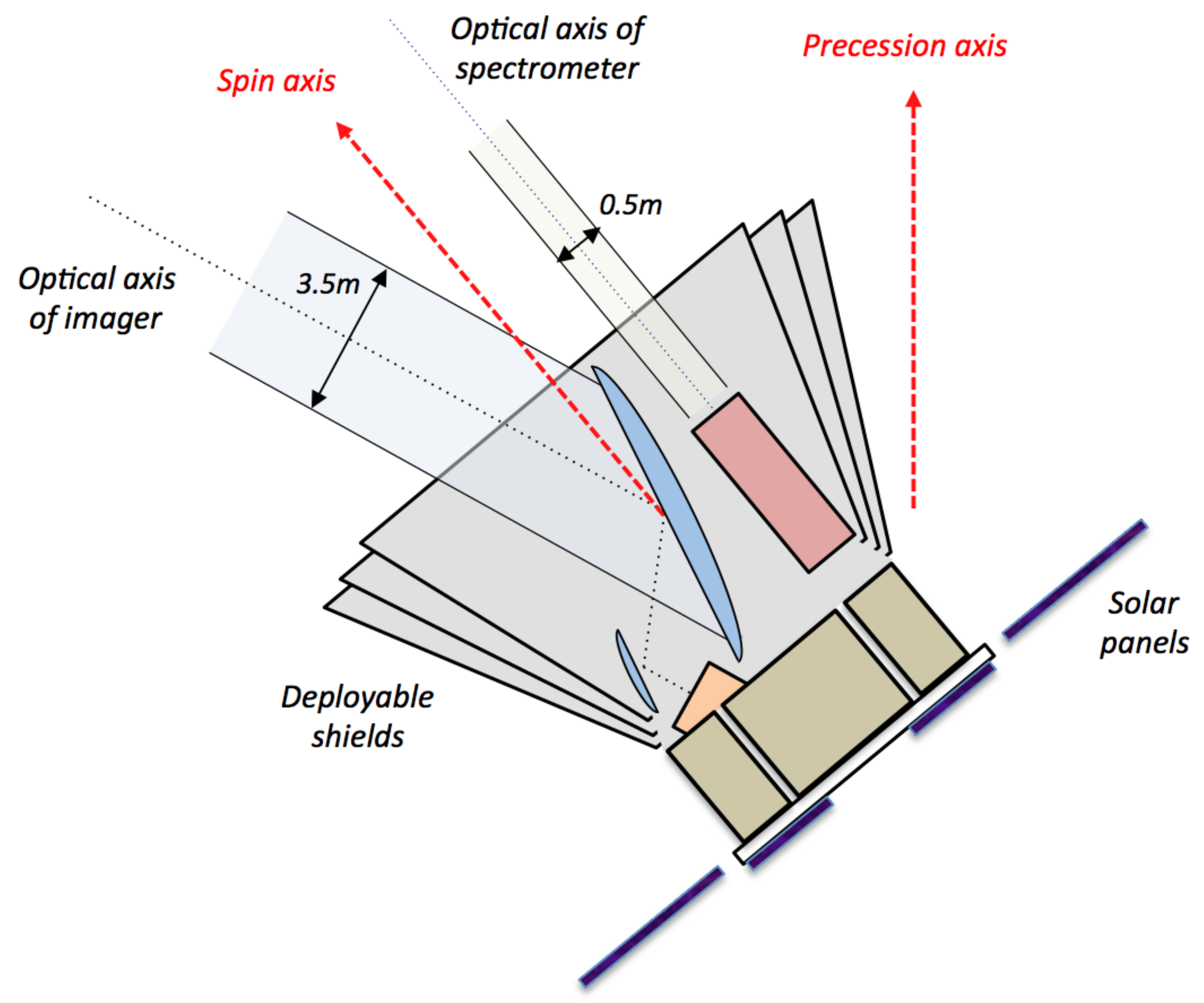}}
\vskip -0.4cm
\caption{{\footnotesize The PRISM spacecraft with its two instruments: PIM, with a 3.5-diameter telescope with a FOV at $\sim$30$^\circ$ from the spacecraft spin axis, and ASP, aligned with the spin axis.}}
\label{fig:compared_instr}
\end{wrapfigure}

The science program 
above requires measuring the sky brightness and polarization at high angular resolution and in 
many frequency bands across a wide spectral range. It also requires measuring the absolute spectrum 
of the sky background with moderate angular and spectral resolution. As a baseline, we 
propose to perform the best possible spectro-polarimetric sky survey in the 30-6000~GHz 
frequency range with two instruments optimized for best joint performance 
sharing a single platform in orbit around the Sun-Earth L2 Lagrange point:
(1) a \emph{polarimetric imager} (PIM) observing with about 30 broad and 300 narrow spectral bands with a diffraction limited angular resolution and a sensitivity limited by the photon noise of the sky emission itself;
and (2)
an \emph{absolute spectro-photometer} (ASP) that will measure sky emission spectra with a spectral resolution between 500~MHz and 15~GHz  and an angular resolution of about 1.4$^\circ$. 
These complementary instruments will map simultaneously the absolute sky intensity and polarization with high sensitivity and with high spectral or spatial resolution. 
The data from both instruments can be binned (in frequency) and smoothed to obtain matching observations with $\delta \nu/\nu \approx 0.25$ and 1.4$^\circ$ resolution, allowing on-sky inter-calibration on large scales (and hence absolute calibration of the PIM).
%
%The high resolution ASP spectra can be re-binned to obtain 1.4$^\circ$ maps with spectral bands closely matching those of the PIM, while the high resolution intensity maps can be smoothed to closely match the angular resolution of the ASP. 
%This will allow for on-sky cross-calibration (and hence absolute calibration of the PIM) and 
This will also enable
correction of the ASP spectra from foreground contamination using high resolution component maps extracted from PIM data (e.g., 
large clusters $y$-distortion in the ASP data and line emission from emitting regions unresolved in the coarse resolution ASP maps).

As the scientific outcome of this mission depends on the complementarity of both instruments and on the control of systematic errors, 
a careful optimization of the ASP and the PIM (number and bandwidth of spectral bands vs. sensitivity) and of the mission (scanning strategy, joint analysis tools) with comprehensive simulations is an essential future phase of the mission study. 

The focal planes of both instruments will be cooled to 0.1K using a cryogenic system adapted from that of \planck, with continuous recycling of the gases for an improved mission duration of 4 years (baseline) or longer. 

\subsection{Instruments}

\vspace{-2mm}
\input polar_imager.tex

\input spectrophotometer.tex

\subsection{Scan strategy}
\input scan_strategy.tex

\subsection{Experimental challenges}
\vspace{-2mm}

\input{experimental_challenges.tex}

\subsection{Ancillary spacecraft}
\input ancillary_spacecraft.tex

%% file: polar_imager.tex
%%%%%%%%%%%%%%%%%%%%%%%%%%%%%%%%%%%%%%%%%%%%%%%%%%%%%%%%%%%
\begin{table}[h]
%{\small
\begin{center}
{\footnotesize
\begin{tabular}{|c|c|c|c|c|c|c|c|c|r|}
\hline 
\hline 
$\nu_0$ 		& Range 		& $\Delta \nu/\nu$ 	& $n_{det}$ 	& $\theta _{\rm fwhm}$ 	& \multicolumn{2}{|c|}{$\sigma_I$ per det}            		&  \multicolumn{2}{|c|}{$\sigma_{(Q,U)}$ per det} 	& Main molec. \& atomic lines \\       
 			&          		&           			&         		& 					& \multicolumn{2}{|c|}{\@ 1 arcmin} 	&  \multicolumn{2}{|c|}{\@ 1 arcmin} 	&  \\       
\cline{6-9}
$\rm GHz$ 	& $\rm GHz$ 	& 	&  			& 		   			& $\mu$K$_{\rm RJ}$    	& $\mu$K$_{\rm CMB}$   		& $\mu$K$_{\rm RJ}$    	& $\mu$K$_{\rm CMB}$  			& \\
\hline 
      30		& 26-34		& .25			&  50			& 17'					& 61.9			& 63.4				& 87.6			& 89.7					& \\  
      36		& 31-41		& .25			&  100		& 14'					& 57.8			& 59.7				& 81.7			& 84.5					& \\  
      43		& 38-48		& .25			&  100		& 12'					& 53.9			& 56.5				& 76.2			& 79.9					& \\  
      51		& 45-59		& .25			&  150		& 10'					& 50.2			& 53.7				& 71.0			& 75.9					& \\  
      62		& 54-70		& .25			&  150		& 8.2'				& 46.1			& 50.8				& 65.2			& 71.9					& \\  
      75   		& 65-85 		& .25			&  150     		&  6.8'				& 42.0			& 48.5				& 59.4   			& 68.6	     				& \\   
      90   		& 78-100 		& .25	 		&  200     		&  5.7'				& 38.0 			& 46.7				& 53.8  			& 66.0     					& HCN \& HCO$^+$ at 89 GHz \\   
      105   		& 95-120		& .25	  		&  250     		&  4.8'				& 34.5			& 45.6				& 48.8 			& 64.4     					& CO at 110-115 GHz\\   
      135   		& 120-150	& .25	  		&  300     		&  3.8'				& 28.6 			& 44.9				& 40.4 			& 63.4    					&  \\   
      160   		& 135-175	& .25	  		&  350     		&  3.2'				& 24.4			& 45.5				& 34.5 			& 64.3    					&  \\   
      185   		& 165-210	& .25	  		&  350     		&  2.8'				& 20.8			& 47.1				& 29.4 			& 66.6    					& HCN \& HCO$^+$ at 177 GHz \\   
      200   		& 180-220	& .20	  		&  350     		&  2.5'				& 18.9			& 48.5				& 26.7 			& 68.6    					&  \\   
      220   		& 195-250	& .25	  		&  350     		&  2.3'				& 16.5			& 50.9				& 23.4 			& 71.9    					& CO at 220-230 GHz \\   
      265   		& 235-300	& .25	  		&  350     		&  1.9'				& 12.2			& 58.5				& 17.3 			& 82.8    					& HCN \& HCO$^+$ at 266 GHz \\   
      300  		& 270-330	& .20	  		&  350     		&  1.7'				& 9.6				& 67.1				& 13.6 			&   94.9    					&  \\   
      320   		& 280-360	& .25	  		&  350     		&  1.6'				& 8.4				& 73.2				& 11.8 			&   103    					& CO, HCN \& HCO$^+$ \\   
      395   		& 360-435	& .20	  		&  350     		&  1.3'				&  4.9			& 107				&  7.0 			&   151    					&  \\   
      460   		& 405-520	& .25	  		&  350     		&  1.1'				&  3.1			& 156				&  4.4 			&   221    					& CO, HCN \& HCO$^+$ \\   
      555   		& 485-625	& .25	  		&  300     		&  55"				&  1.6			& 297				&  2.3 			&    420					&  C-I, HCN, HCO$^+$, H$_2$O, CO \\   
      660   		& 580-750	& .25	  		&  300     		&  46"				&  0.85			& 700				&  1.2 			&    990   					&  CO, HCN \& HCO$^+$ \\   
\hline 
\hline 
\cline{6-9}
			& 			& 	&  			& 		   			& nK$_{\rm RJ}$    	& kJy/sr   						& nK$_{RJ}$    	& kJy/sr    					&  %(many in all channels below)
			\\
\hline 
      800   		& 700-900 	& .25			& 200      		&  38"				& 483  		& 9.5  						& 683   		& 13.4      						&  \\   
      960   		& 840-1080	& .25	 		& 200      		&  32"				& 390 		& 11.0						&  552  		&  15.6     						&  \\   
      1150   		& 1000-1300	& .25	  		& 200      		&  27"				& 361 		& 14.6 						&  510  		&   20.7    						&  \\   
      1380   		& 1200-1550	& .25	  		& 200      		&  22"				& 331 		& 19.4						&  468  		&    27.4   						&  N-II at 1461 GHz\\   
      1660   		& 1470-1860	& .25	  		& 200     		&  18"				& 290 		& 24.5 						&  410  		&    34.7   						&  \\   
      1990   		& 1740-2240	& .25	  		& 200      		&  15"				& 241 		&  29.3 						&  341  		&    41.5   						&  C-II at 1900 GHz\\   
      2400   		& 2100-2700	& .25  		& 200      		&  13"				& 188		&  33.3 						&  266  		&    47.1   						&  N-II at 2460 GHz\\   
      2850   		& 2500-3200	& .25	  		& 200      		&  11"				& 146		&  36.4 						&   206 		&    51.4   						&  \\   
      3450   		& 3000-3900	& .25	  		& 200      		&  8.8"				& 113 		&  41.4 						&   160 		&    58.5   						&  O-III at 3393 GHz\\   
      4100  		& 3600-4600	& .25	  		& 200      		&  7.4"				& 98 			&  50.8 						&   139 		&    71.8   						&  \\   
      5000   		& 4350-5550	& .25	  		& 200      		&  6.1"				& 91 			&  70.1 						&   129 		&     99.1  						&  O-I at 4765 GHz\\   
      6000   		& 5200-6800	& .25  		& 200      		&  5.1"				& 87	 		&  96.7 						&    124		&     136  						&  O-III at 5786 GHz\\   
\hline 
\hline 
\end{tabular}
}
\end{center}
\vspace{-4mm}
\caption{{\footnotesize The 32 broad-band channels of the polarized imager
with a total of 7600 detectors. Sensitivities are averages for sky regions 
at galactic latitude and ecliptic latitude both higher than $30^\circ$. 
A detector noise level equal to the sky photon noise is assumed. The mission 
sensitivity per frequency channel is the sensitivity per detector divided by $\sqrt{n_{det}}$. }}
\label{tab:PIM-bands}
\end{table}
%%%%%%%%%%%%%%%%%%%%%%%%%%%%%%%%%%%%%%%%%%%%%%%%%%%%%%%%%%%

%%%%%%%%%%%%%%%%%%%%%%%%%%%%%%%%%%%%%%%%%%%%%%%%%%%%%%%%%%%
%\begin{figure}[!h]
%\begin{center}
%    \begin{tabular}{lr}
%    \vspace{5mm}
%    %{\includegraphics[trim=0cm 0cm 0cm 0cm, clip=true,width=0.45\textwidth]{Figures/sensitivity_comparison.ps}}  \hspace{5mm}
%    {\includegraphics[trim=0cm 0cm 0cm 0cm, clip=true,width=0.46\textwidth]{Figures/spacecraft.pdf}}  \hspace{2mm}
%    {\includegraphics[trim=0cm 0cm 0cm 0cm, clip=true,width=0.49\textwidth]{Figures/payload-design.pdf}}  \\
%    \end{tabular}
%    \caption{Left: Spacecraft (rough sketch). Right: schematic of a possible payload \textcolor{red}{[TO BE REPLACED]}.}
%    \label{fig:compared_instr}
%\end{center}
%\end{figure}

%%%%%%%%%%%%%%%%%%%%%%%%%%%%%%%%%%%%%%%%%%%%%%%%%%%%%%%%%%%

\mypar{The polarimetric imager:}
The optical configuration relies on a dual off-axis mirror telescope with a 3.5$\,$m projected aperture diameter primary and a 0.8$\,$m diameter secondary coupled to a multi spectral band polarimeter. 
The broad-band PIM comprises 32 main channels of $\delta \nu/\nu \approx .25$ relying on dual-polarized 
pixel arrays (Table~\ref{tab:PIM-bands}). At frequencies below 700 GHz, the emphasis 
is on the sensitivity and control of systematics for CMB and SZ science.

The whole frequency range will also be covered at higher spectral resolution ($\delta \nu/\nu 
\approx  .025$) to map spectral lines. The $\sim$300 frequency channels (not listed in Table~\ref{tab:PIM-bands}) 
will be obtained using antenna coupled bolometers and channelizers to split the spectral 
band of each broad-band horn into 5-10 narrow frequency bands, with similar
numbers of narrow-band and broad-band detectors.
The sensitivity to continuum emission per detector is reduced in the narrow-band channels 
as compared to the broad-band channels, but the sensitivity to spectral lines is better by a factor of about 2-3.

%% file: spectrophotometer.tex
\mypar{The absolute spectrophotometer: }
A Martin-Puplett Fourier Transform Spectrometer (FTS) will allow for a large throughput and
sensitivity, differential measurements (the sky is compared to an
internal blackbody calibrator as in COBE-FIRAS), and a variable
spectral resolution. Dichroics at the two output ports
can optionally
split the full 30 - 6000~GHz range into sub-bands with reduced 
photon noise. The instrument is cooled at 2.7K, so that the 
bolometric detector sensitivity is limited by photon noise from the sky.
Two operating modes are available:
high-resolution ($\Delta \nu \sim 0.5\,$GHz) and low-resolution
($\Delta \nu \sim 15\,$GHz). The sensitivity of the high-resolution
mode is 30 times worse than for the low-resolution mode. The
instrument beam is aligned with the spin axis of the satellite, so
that precession has a negligible effect during the
interferogram scan ($\sim$1s/10s long in the low-res/high-res
mode). The main characteristics for three possible configurations
of the instrument are detailed in Table \ref{tab:spectrometer}.

\begin{table}[h]
\begin{center}
{\footnotesize
\begin{tabular}{|c|c|c|c|c|c|}
\hline 
\hline Band & Resolution & $A\Omega$ & Background & NEP$\nu$ & Global 4-yr mission 
\\
(GHz) & (GHz) & (cm$^2$sr) & (pW) &
(W/m$^2$/sr/Hz$\times\sqrt{\rm s}$) & sensitivity (W/m$^2$/sr/Hz)
\\
\hline \hline
30-6000 & 15 & 1 & 150 & $1.8\times 10^{-22}$ & $1.8 \times 10^{-26}$ \\
\hline
30-500 & 15 & 1 & 97 & $7.0\times 10^{-23}$ & $7.2 \times 10^{-27}$ \\
500 - 6000 & 15 & 1 & 70 & $1.7\times 10^{-22}$ & $1.7 \times 10^{-26}$ \\
\hline
30-180 & 15 & 1 & 42 & $3.5\times 10^{-23}$ & $3.6 \times 10^{-27}$ \\
180-600 & 15 & 1  & 57 &  $6.3\times 10^{-23}$ & $6.5 \times 10^{-27}$ \\
600-3000 & 15 & 1 & 20 &  $7.4\times 10^{-23}$ & $7.6 \times 10^{-27}$ \\
3000-6000 & 15 & 1 & 28 &  $1.6\times 10^{-22}$ & $1.6 \times 10^{-26}$ \\
\hline
\hline 
\end{tabular}
}
\end{center}
%\\
\vspace{-4mm}
\caption{{\footnotesize FTS performance of
three possible configurations for photon noise limited
detectors.
%, operating in a radiative background comprising
%CMB, high galactic latitude interstellar dust, CIB, and high ecliptic latitude
%interplanetary dust emission. 
With an entrance pupil 50
cm in diameter, the baseline throughput is $\sim 1 \, {\rm cm}^2{\rm sr}$ and
the angular resolution 1.4$^\circ$. The theoretical monopole sensitivity
for each spectral bin is reported in the last column assuming 4 years of observation and 75\% useful sky. The actual
sensitivity, taking into account efficiency factors can be 2-3 times worse. 
Line 1 is a configuration with an ultra-wide spectral coverage obtained
with one detector in both output ports. In lines 2-3 the detectors 
at the output ports are sensitive to different bands. 
%
%Similar numbers apply if each output port is split in two sub-bands 
%with identical dichroics on both outputs. 
In lines 4-7 each output port is
split into two sub-bands using dichroics to minimize photon noise in
the low-frequency bins.}}
\label{tab:spectrometer}
\end{table}

Using detectors with $A\Omega \! \sim \! 1 \, {\rm cm}^2{\rm sr}$ and
angular resolution $\sim$1.4$^\circ$, we estimate that the CIB
can be measured with $S/N=10$ in a fraction of a second at 1500 GHz 
and in $\sim 10$ seconds 
at 140 GHz, while a $y$-distortion $\sim 10^{-8}$ can be measured
with $S/N=10$ at 350$\,$GHz in two hours of integration. Recombination
lines could be measured integrating over the whole mission if the
overall stability of the instrument and the quality of the
reference blackbody are sufficient.

The main issue for this instrument is the control of systematic
effects. The instrument design allows for a number of zero tests
and cross-checks on the data. The main problem is to control the
blackness of the reference and calibration blackbodies with the
required accuracy. Reflectivities lower than $R=-50 / \! -\! 60\,$dB
have been obtained in the frequency range of interest in the
Planck and ARCADE references. We plan to achieve $R<-70\,$dB
building on these experiences through a combination of
electromagnetic simulations and laboratory emissivity measurements
on improved shapes and space-qualified materials.

%% file: scan_strategy.tex
The observing strategy must provide: 
(1) full sky coverage for both instruments; 
(2) cross-linked scan paths and observation of all sky pixels in many orientations for all 
detectors of the PIM;
(3) fast scanning of the PIM to avoid low-frequency drifts; 
(4) slow scanning for the ASP field of view to allow for few seconds 
long interferogram scans with negligible depointing; 
(5) avoiding direct solar radiation on the payload.
These requirements can be satisfied by a spinning spacecraft with the ASP 
aligned along the spin axis and the PIM offset by $\theta_{\rm spin} 
\approx 30^\circ$ (Fig. \ref{fig:compared_instr}). During each spacecraft rotation 
(with  $\omega_{\rm spin}$ of a few rpm), the PIM scans 
circles of diameter $\approx 2\theta_{\rm spin}$ while the APS rotates in place.
A slow precession of the spin axis (with a period between a few hours and one 
day) with a precession angle $\theta_{\rm prec} \approx 45^\circ$ results in slow scans of the 
ASP on large circles of diameter $2\theta_{\rm prec}$. Finally, the precession axis 
evolves by about 1$^\circ $ per day along the ecliptic plane to keep the 
payload away from the Sun, and also slowly moves perpendicular to the ecliptic plane 
so as to map the ecliptic poles. Deployable screens isolate the payload from the 
heat from the Sun, providing a first stage of passive cooling 
to $\approx 40\,$K.

%% file: experimental_challenges.tex
\mypar{Telescope temperature:} 
Actively cooling the telescope to 4$\,$K (mission objective)
instead of 40$\,$K (achievable by passive cooling)
substantialy improves the sensitivity, especially for 
frequencies above 200~GHz.
\mission\ will benefit 
from the development activities for the SPICA mission, the telescope of which is based on a 
3.5~m diameter primary cooled to 5~K.

\mypar{Polarization modulation:} The baseline, similar to the solution proposed in the 
previous SAMPAN and EPIC studies, relies on the scanning strategy and the rotation of the 
entire payload. However alternative strategies such as the use of a half-wave plate in front of 
the focal plane (the receivers being the major source of instrumental polarization) could be 
considered during a trade-off analysis.

\mypar{Detectors:} Direct detectors (such as TES bolometers, CEBs or KIDs) are the most 
sensitive detectors at mm wavelengths. Bolometers have achieved photon noise limited in-flight 
performance with the Planck \citep{2011A&A...536A...4P} and Herschel 
\citep{2010A&A...518L...3G} missions. Large bolometer arrays with thousands of pixels are 
currently used on large ground-based telescopes. They are currently not proven as a viable 
technology for 30 to 70~GHz but it is likely that their efficiency will improve in the next 
few years at low frequencies. For instance studies \citep{2007stt..conf...93K} have shown that 
70$\,$GHz CEBs could lead to NEPs of (few)$\times 10^{-18}~W\cdot Hz^{1/2}$. 
As an alternative solution, 
the \mission\ instruments could take advantage of the recent breakthroughs in 
cryogenic HEMT technology, with sensitivities predicted to reach 2-3 times the quantum limit 
up to 150-200$\,$GHz (instead of 4-5 times up to 100$\,$GHz so far). In addition, these 
devices allowing for cryogenically cooled miniaturized polarimeter designs will simplify the
thermo-mechanical design. Hence, while a single detector technology throughout the instruments 
would be preferable, the option of using a combination of HEMTs and bolometers remains open 
(Table \ref{tab:techno}).

\begin{table}[t]
\centering{}%
{\footnotesize
\begin{tabular}{|c|c|c|c|c|c|c|}
\hline 
$\nu_{c}$ ange & Req. NEP & Req. $\tau$ & \multicolumn{4}{c|}{Focal plane technology}\tabularnewline
\hline 
\multirow{2}{*}{$\left[GHz\right]$} & \multirow{2}{*}{$\left[10^{-18}\, W/\sqrt{Hz}\right]$} 
& \multirow{2}{*}{$\left[ms\right]$} & \multicolumn{2}{c|}{Detector technology} & \multicolumn{2}{c|}{Optical coupling}\tabularnewline
\cline{4-7} 
 &  &  & Baseline & Backup & Baseline & Backup\tabularnewline
\hline 
\hline 
{\small 30-75} & 3.3 -- 5.7 & {\small 2.96 -- 1.18} & {\small TES} & {\small HEMT} & {\small MPA/CSA} & {\small HA}\tabularnewline
\hline 
{\small 90-320} & 4.6 -- 7 & {\small 1.18 -- 0.4} & {\small TES} & {\small KIDS} & {\small HA+POMT} & {\small MPA}\tabularnewline
\hline 
{\small 395-660} & 0.94 -- 3.1 & {\small 0.4 -- 0.13} & {\small TES} & {\small KIDS} & {\small MPA/CSA} & {\small LHA}\tabularnewline
\hline 
{\small 800-6000} & 0.011 -- 0.63 & {\small 0.13 -- 0.01} & {\small KIDS} & {\small HEB/CEB} & {\small MPA/CSA} & {\small LHA}\tabularnewline
\hline 
\multicolumn{1}{c}{} & \multicolumn{1}{c}{} & \multicolumn{1}{c}{} & \multicolumn{1}{c}{} & \multicolumn{1}{c}{} & \multicolumn{1}{c}{} & \multicolumn{1}{c}{}\tabularnewline
\end{tabular}
}
\vspace{-4mm}
\caption{\footnotesize Required NEP and time constants for various frequency ranges and 
corresponding baseline and backup focal plane technology. TES: Transition Edge Sensors 
(Technology Readiness Level 5); HEMT: High Electron Mobility Transistor (TRL 5); KID: Kinetic 
Inductance Detector (TRL 5); HEB: Hot Electron Bolometer (TRL 4); CEB: Cold Electron Bolometer 
(TRL 3); HA: Horn Array (TRL 9); LHA: Lithographed Horn Array (TRL 5); MPA: Multichroic Planar 
Antenna (TRL 4); CSA: Crossed Slot Antenna (TRL 5); POMT: Planar Ortho-Mode Transducer (TRL 
5)}
\label{tab:techno}
\end{table}

\mypar{Detector time constants:} The fast scanning of the \mission\ mission requires fast 
detector time constants, of order 1$\,$ms at 100 GHz, down to $\sim 10\, \mu$s at 6 THz. These 
time constants are challenging (especially at high frequencies), but have already been 
achieved with recent TESs, KIDs or CEBs.

%% file: ancillary_spacecraft.tex
We propose that the mission include a small ancillary spacecraft serving the 
following functions:

\mypar{Telecommunication:} The high resolution mapping of the full sky with the many 
detectors of \mission\ with a lossless compression of 4 gives a total data rate of 
$\sim350\,$Mbit/s (of which $300\,$Mbit/s is from the channels above 700 GHz). Further 
on-board reduction by a factor $\sim\!10\!-\!20$ can be achieved by averaging the timelines 
of detectors following each other on the same scan path (after automatic removal of spikes 
due to cosmic rays) to yield a total data rate $<40\,$Mbit/s (a few times greater than Euclid or 
Gaia). A phased-array antenna or counter-rotating antenna on the main spacecraft is an option.
Decoupling the communication function from the main spacecraft using an 
ancillary spacecraft
as 
an intermediate station for data transmission will allow for a maximally flexible scanning 
strategy for the best polarization modulation and full sky coverage.

\mypar{In-flight calibration:} The hardest \mission\ design problem is ensuring that 
the performance is limited by detector noise rather than systematic effects and calibration 
uncertainties. While pre-flight calibration is necessary, an 
ancillary spacecraft
fitted with calibrated, 
polarized sources could be used for precise in-flight calibration of the polarization 
response and polarization angles of the detectors, and for main beams and far sidelobe 
measurements down to extremely low levels (below -140~dB) at several times during the 
mission lifetime.

%% file: relation_to_other_initiatives.tex
%\section{Complementarity and competition with other initiatives}

\mypar{B-mode experiments:}
Searching for primordial gravitational waves through B-mode
polarization is the principal science driver of numerous
suborbital experiments
(e.g., {\it BICEP, QUIET, SPIDER, ACTPol, SPTPol, QUBIC, EBEX, PolarBear, QUIJOTE})
despite considerable limitations
due to atmospheric opacity, far-side lobe pickup from the ground, and 
unstable observing conditions that make controlling systemic errors especially
difficult, particularly
on the largest angular scales where
the B mode signal is largest. Forecasts of $r$ from ground-based experiments
are often impressive but assume very simple foregrounds. For this reason a detection of $r$
from the ground would provide a strong motivation for a confirmation and more precise
characterization from space.
Moreover, two US space missions concepts, {\it CMBPol}
and {\it PIXIE,} and one in Japan, {\it LiteBird,} have been proposed, but
none has yet been funded. 
Among the current space mission concepts, 
\mission\ is the most ambitious 
and encompasses the broadest science case.
{\it LiteBird} is a highly-targeted, low-cost Japanese B-mode
mission concept, in many respects similar to the {\it BPol} mission
proposed to ESA in 2007. With its coarse angular resolution and
limited sensitivity, {\it LiteBird} would be able to detect B-modes
assuming that $r$ is not too small and that the foregrounds are
not too complicated. {\it LiteBird,} however, lacks the angular resolution
needed to make significant contributions to other key science 
objectives. The US {\it EPIC-CS} mission  
is the most similar to the present proposal but has considerably 
less frequency coverage, fewer frequency bands, and no
absolute spectral capability. The US mission concept
{\it PIXIE} proposes an improved version of the {\it FIRAS} spectrometer
to measure B-modes 
and perform absolute spectroscopy simultaneously, but with an effective 
resolution of only $2.6^\circ .$ 

\mypar{Cluster observations:}
When \mission\ flies, the {\it eROSITA} X-ray survey will likely be the only
deeper all-sky cluster survey available.
20--30 times more sensitive than ROSAT,
{\it eROSITA}'s principal goal is to explore cosmological models using galaxy clusters.
Forecasts predict that {\it eROSITA} will detect $\sim10^5$ clusters at more than
100 photon counts, which is sufficient to provide a good detection and in many cases to
detect the source as extended in X-rays. The main survey provides a good sample of galaxy
clusters typically out to $z=1$ with some very massive and exceptional clusters at larger
distance.

The large majority of these clusters will be re-detected by \mission\ and thus
provide an invaluable inter-calibration of X-ray and SZ effect cluster cosmology, provide
determinations of cluster temperatures by combining the two detection techniques, and obtain
independent cluster distances for many thousands of clusters whose X-ray temperatures and
shape parameters can be obtained from the X-ray survey.
With $\sim10^6$ clusters detected with \mission, one can further exploit the
{\it eROSITA} survey data by stacking in a way similar to the analysis of the 
X-ray signals from 
the {\it ROSAT} All-Sky Survey for {\it SDSS} detected
clusters (Rykoff et al. 2008).

\mypar{Other sub-millimeter/far-infrared initiatives:}
Existing ({\it APEX, ASTE, IRAM 30m, LMT}) and future ({\it CCAT}) ground-based
single-dish submillimeter observatories are not as sensitive above 300 GHz
as \mission , mainly because of the limitations of
observing through the atmosphere.  
Interferometers ({\it ALMA, CARMA, PdB Interferometer, SMA}) are ill-suited
to observing large fields.  Moreover most interferometers are
insensitive to
large-scale structure. {\it SKA} will span the radio range from 0.07 GHz up to 20 
GHz, and will be the perfect complement to \mission, with more than $10^{9}(f_{sky}/0.5)$ 
HI galaxies in a redshift range $0 < z < 1.5$, and maps of the epoch of reionization above $z\sim 6$.

\mission\  will map the full-sky, large-scale continuum emission at higher
sensitivities than ground based single-dish telescopes.  Bright compact
sources found by \mission\ in its all-sky surveys can subsequently be
observed in more detail by interferometers.  Observations 
can be combined to produce superior maps of selected sky regions.
The Atacama Large Millimeter Array ({\it ALMA}), operating in the 
range 30-1000 GHz, will complement \mission\ with follow-up of sources 
and clusters detected by \mission,
mapping their structure in total intensity, polarization and spectral line
at high angular and spectral resolution.

{\it CCAT} will initially have two imaging instruments.  At low frequencies, {\it LWCam} on {\it CCAT} will 
be able to detect sources below the \mission\ confusion limit relatively quickly. 
However variations in atmospheric transmissivity and 
thermal radiation from the atmosphere make it difficult for {\it CCAT} to map large scale 
structures. At high frequencies, {\it SWCam} will have difficulty mapping large areas to the 
confusion limit of \mission.  Based on the specifications from Stacey et al. (2013), {\it CCAT} can 
map an area of 1 square degree at 857 GHz to a sensitivity of 6 mJy (the \mission\ confusion limit) 
within 1 hour.  To map the entire southern sky to this same depth 
requires $\sim900$ days (24h) with optimal observing conditions.  Such large scale 
observations will not be feasible with {\it CCAT.} \mission\ is needed to produce all-sky 
maps in these frequency bands. 
\mission\ will produce maps at the same resolution as {\it Herschel.}  However {\it Herschel} was 
able to map only a limited portion of the sky. 

Few previous infrared telescopes have performed all-sky surveys in the
bands covered by \mission. {\it Akari} was the last telescope to perform
such observations, 
but the data is at much lower sensitivity and resolution
and is not yet publicly available.
Several other prior telescopes {\it (Spitzer, Herschel)} as well as
the airborne observatory {\it SOFIA} and the future mission {\it SPICA} have observed or will 
observe in the 600-4000 GHz range, but only
over very limited areas of the sky. Furthermore, except for a
few deep fields, they observe objects already identified in other
bands.  \mission\ will be able to perform observations with sensitivities
comparable to {\it Herschel} or better, but covering the entire sky in many frequency bands.